\documentclass[aps,twocolumn,prd,superscriptaddress,showpacs,preprintnumbers,nofootinbib]{revtex4-1}

\usepackage[latin9]{inputenc}

\usepackage{mathtools}
\usepackage{amsfonts}
\usepackage{mathrsfs}
\usepackage{bbm}
\usepackage{slashed}

\usepackage{graphicx}
\usepackage{color}

\usepackage{placeins}
\usepackage{booktabs}
\usepackage[caption=false]{subfig}

\usepackage{xspace}
\usepackage{siunitx}
\usepackage{hyperref}

\setkeys{Gin}{width=0.48\textwidth}

\graphicspath{
	{./figures/}
	{../../talks+proceedings/general_figures/}
}

\def\sBRST{\mathfrak s}

\def\s0#1#2{\mbox{\small{$ \frac{#1}{#2} $}}}
\def\0#1#2{\frac{#1}{#2}}

\def\flavT#1{T^{#1}_{\rm f}}
\def\colT#1{T^{#1}_{\rm c}}
\def\dd{\text{d}}

\newcommand{\imag}{\text{i}}
\newcommand{\Tr}{{\text{Tr}}}

\newcommand{\ltpre}[1][0pt]{\mathrel{\raisebox{#1}{\scriptsize$\parallel$}}}
\newcommand{\lt}{{\ltpre[1.2pt]}}
\newcommand{\trans}{ \bot}
\newcommand{\LambdaSTI}{\Lambda_{\text{\tiny{STI}}}}
\newcommand{\rebosA}{A_{\bar q \flavT{a} q,k}}
\newcommand{\rebosB}{B_{\bar q \flavT{a} q,k}}
\newcommand{\identityflavour}{\mathbbm{1}_{\rm f}}
\newcommand{\identitycolour}{\mathbbm{1}_{\rm c}}
\newcommand{\identityspinor}{\mathbbm{1}_{\rm s}}
\newcommand{\psym}{\bar{p}}

\newcommand{\eqnewline}{\nonumber\\[2ex]}
\newcommand{\myhfill}{\hfill\textcolor{white}{.}}
\newcommand*{\eg}{e.g.\@\xspace}
\newcommand*{\ie}{i.e.\@\xspace}

\newcommand*{\cf}{cf.\@\xspace}

\def\eq#1{\eqref{#1}}
\def\Fig#1{Fig.~\ref{#1}}
\def\Figs#1#2{Fig.~\ref{#1} and \ref{#2}}
\def\Tab#1{Tab.~\ref{#1}}
\def\Sec#1{Sec.~\ref{#1}}
\def\Secs#1#2{Sec.~\ref{#1} and \ref{#2}}
\def\App#1{App.~\ref{#1}}

\newcommand{\gettitle}{Non-perturbative 
quark, gluon and meson correlators of unquenched QCD}

\hypersetup{
	pdftitle={\gettitle},
	pdfauthor={Cyrol, Mitter, Pawlowski, Strodthoff},
	pdfkeywords={QCD} {Functional Renormalisation Group} {Correlation Functions} {Non-perturbative} {Unquenched},
	bookmarksopen=false,
	bookmarksnumbered=true
}

\begin{document}

\title{\gettitle}

\author{Anton K. Cyrol}
\affiliation{Institut f\"ur Theoretische Physik,
	Universit\"at Heidelberg, Philosophenweg 16,
	69120 Heidelberg, Germany
}

\author{Mario Mitter}
\affiliation{Institute of Physics,
	NAWI Graz, University of Graz, Mozartgasse 14,
	8010 Graz, Austria
}

\author{Jan M. Pawlowski} 
\affiliation{Institut f\"ur Theoretische Physik,
	Universit\"at Heidelberg, Philosophenweg 16,
	69120 Heidelberg, Germany
}
\affiliation{ExtreMe Matter Institute EMMI,
	GSI, Planckstr. 1,
	64291 Darmstadt, Germany
}

\author{Nils Strodthoff}
\affiliation{Nuclear Science Division,
	Lawrence Berkeley National Laboratory,
	Berkeley, CA 94720, USA
}


\begin{abstract}
  We present non-perturbative first-principle results for quark-,
  gluon- and meson $1$PI correlation functions of two-flavour
  Landau-gauge QCD in the vacuum. These correlation functions carry
  the full information about the theory. They are obtained by solving
  their Functional Renormalisation Group equations in a systematic
  vertex expansion, aiming at apparent convergence. This work
  represents a crucial prerequisite for quantitative first-principle
  studies of the QCD phase diagram and the hadron spectrum within this
  framework.

  In particular, we have computed the gluon, ghost, quark and
  scalar-pseudoscalar meson propagators, as well as gluon,
  ghost-gluon, quark-gluon, quark, quark-meson, and meson
  interactions. Our results stress the crucial importance of the
  quantitatively correct running of different vertices in the
  semi-perturbative regime for describing the 
  phenomena and scales of confinement and spontaneous chiral symmetry
  breaking without phenomenological input.
\end{abstract}

\maketitle

\section{Introduction} 

Over the past decades, the resolution of the QCD phase structure as
well as the hadron spectrum has been at the forefront of research in
theoretical hadron physics. The most important open questions include
the existence and location of a critical point in the QCD phase
diagram, the spectrum of higher hadronic resonances, as well as the
computation of the dynamical properties of QCD matter from its
microscopic description. Answering these qualitative and quantitative
questions requires controlled first-principle approaches. However,  
lattice as well as functional approaches face
various conceptual as well as numerical challenges. These range from
the sign problem in the former, and the problem of convergent
expansion schemes in the latter, to the need for real-time numerical 
methods for the dynamics of quantum systems and the hadron
spectrum. Thus, complementary and combined studies within different
approaches offer important cross checks as well as the potential to
overcome problems that cannot be addressed within one method alone.

Recently, the focus has shifted towards top-down approaches
\cite{Mitter:2014wpa,Cyrol:2016tym,Binosi:2014aea,Williams:2015cvx,Huber:2016tvc}
within the functional continuum methods, see \eg
\cite{Litim:1998nf,Berges:2000ew,Pawlowski:2005xe,Gies:2006wv,
Schaefer:2006sr,Rosten:2010vm,Braun:2011pp,vonSmekal:2012vx,Alkofer:2000wg,
Roberts:2000aa,Fischer:2006ub,Fischer:2003rp,Aguilar:2008xm,Fischer:2008uz,Binosi:2009qm,
Maas:2011se,Boucaud:2011ug,Eichmann:2016yit,Aguilar:2012rz,Ayala:2012pb,Pelaez:2013cpa,Blum:2014gna,
Eichmann:2014xya,Gracey:2014mpa,Gracey:2014ola,Binosi:2014kka,Cyrol:2014kca,Braun:2014ata,
Rennecke:2015eba,Boucaud:2017obn,Bermudez:2017bpx,Gracey:2017yfi,Reinosa:2017qtf,
Pelaez:2017bhh,Braun:2017srn,Contant:2017gtz}.
Within a top-down approach, only the fundamental parameters of QCD
enter as input and no phenomenological modelling is required.  Such a
framework has recently been established by the fQCD collaboration
\cite{fQCD:2016-10}. This collaborative effort was initiated with the
goal of using the Functional Renormalisation Group (FRG) as a
quantitative first-principle approach to continuum QCD. Primary
applications of this correlation-function-based approach are the QCD
phase structure and the hadron spectrum, as well as real-time dynamics
of QCD. In \cite{Mitter:2014wpa} and \cite{Cyrol:2016tym} we have
presented quantitative results for the correlators of quenched
two-flavour Landau-gauge QCD and Landau-gauge Yang-Mills theory. The
advances established by these two works build the foundation for the
present study of the coupled unquenched system of equations for the
QCD correlation functions. This work constitutes a crucial
prerequisite for future quantitative first-principle studies at finite
temperature and finite chemical potential.
This is confirmed by studies of low-energy effective models which show 
that mismatches in the fluctuation scales inevitably 
cause large systematic errors \cite{Helmboldt:2014iya}.
Furthermore, this top-down
approach allows the formulation of QCD-enhanced effective models for
different aspects of the strong interaction under extreme conditions,
see \eg \cite{Haas:2013qwp,Herbst:2013ufa,Fu:2015naa,Heller:2015box,Fu:2016tey} 
for corresponding studies on the equation of state and the axial
anomaly at finite temperatures.

We present a computationally sophisticated analysis of the
self-consistently coupled unquenched system of FRG equations for QCD
correlation functions.  This is not a means in itself, but is
necessitated by the mechanism of spontaneous chiral symmetry breaking.
We found already in earlier studies that even small deviations in the
running couplings in the semi-perturbative regime can lead to the
complete absence of chiral symmetry breaking \cite{Mitter:2014wpa}.
Hence, the consistent semi-perturbative running of different vertices
is of crucial importance. Additionally, a full quantitative resolution
of the quark-gluon interaction turns out to be of qualitative
importance for spontaneous chiral symmetry breaking, \cf
\cite{Alkofer:2008tt,Williams:2014iea,Mitter:2014wpa,Williams:2015cvx,Binosi:2016wcx,Aguilar:2016lbe}.
In order to guarantee the crucial self-consistent running of
  the vertices, we use the Slavnov-Taylor identity (STI) to constrain
  the transverse quark-gluon vertex in the perturbative and
  semi-perturbative regime. In particular, we take into account loop
  corrections to the STI, see \eg
  \cite{Davydychev:2000rt,Aguilar:2010cn,Rojas:2013tza,Aguilar:2014lha,Aguilar:2016lbe}.
In the non-perturbative regime, on the other hand, the STI cannot
constrain the transversely projected vertex.  Consequently, we solve
the full flow equation for the vertex in this regime.

The paper is organised as follows: In \Sec{sec:fqcd} we review the
FRG treatment of continuum QCD in a vertex expansion scheme, discuss
running couplings and STIs as well as the general computational
framework. Results for propagators and vertices are presented and discussed in
\Secs{sec:results}{sec:AppCon}. We summarise and conclude in
\Sec{sec:summary}.
Details on the momentum-dependent generalisation
of the dynamical hadronisation procedure, the truncation scheme, the
interaction vertices, the modified Slavnov-Taylor identities and the
regularisation scheme are provided in appendices.

\section{QCD from the Functional Renormalisation Group} 
\label{sec:fqcd}

In this work we present a self-consistent solution of the system of
Functional Renormalisation Group (FRG) equations for a large subset of
the QCD correlation functions. It builds on previous works, and we
refer to
\cite{Mitter:2014wpa,Braun:2014ata,Rennecke:2015eba,Cyrol:2016tym} for
more technical details. The classical gauge-fixed action in a
covariant gauge is given by
\begin{align}
	S_{\text{\tiny QCD}}=&\,\int_x\,\014 F_{\mu\nu}^a  F_{\mu\nu}^a   +\int_x\left( 
		\lambda^a   \partial_\mu A^a_\mu-\frac{\xi}{2}\lambda_a^2 \right)
	\eqnewline
		& -\int_x\,\bar c^a \partial_\mu D^{ab}_\mu
		c^b+\int_x \bar q\, \slashed{D} \,q \,.
\label{eq:sclassical} 
\end{align} 
Here, $\xi$ denotes the gauge fixing parameter, which is taken to zero
in the Landau gauge and $\int_x=\int \dd^4 x\,$.  In
\eq{eq:sclassical} we have introduced the auxiliary Nakanishi-Laudrup
field $\lambda\,$. It facilitates the discussion of the STIs in
\App{app:quarkghostscatteringkernel}. On its equations of
motion the second term in \eq{eq:sclassical} reduces to the usual
gauge fixing term
\begin{align}
	S_{\rm gf}[A]\Bigr|_{\rm \lambda=\lambda_{\text{\tiny{EoM}}}}= \01{2 \xi} \int_x (\partial_\mu A_\mu^a)^2\,,
\end{align}
and the dependence on $\lambda$ will be suppressed in the remainder of
the work, except in the discussion of the STI in
\App{app:quarkghostscatteringkernel}. The field strength
tensor and the covariant derivative $\partial_\mu - \imag g \,A_\mu$ in 
the adjoint representation are given by
\begin{eqnarray}
	F^a_{\mu\nu} &=&  \partial_\mu A^a_\nu-\partial_\nu A^a_\mu+ 
		g f^{abc}A_\mu^b A_\nu^c\,,
	\eqnewline
	D^{ab}_{\mu} & = &\delta^{ab}\partial_\mu- f^{abc} A_\mu^c \,, 
\end{eqnarray}
where in the adjoint representation we have
$(T^c_{\text{\tiny{ad}}})^{ab} =- \imag f^{ab c}\,$. 
The fundamental generators $\colT{a}$
satisfy the defining Lie algebra commutation relation and are normalised to 
$1/2\,$:
\begin{align}
	\left[\colT{a},\colT{b}\right] = \imag f^{abc}\colT{c}\,,\qquad 
	\Tr\left(\colT{a}\colT{b}\right) = \frac{1}{2}\delta^{ab}\,.
\end{align}
The Dirac operator in \eq{eq:sclassical} in the fundamental representation reads 
\begin{align}
\label{eq:Dirac} 
\slashed{D}  = \gamma_\mu D_\mu \,,\quad {\rm with} \quad D_\mu =  \partial_\mu- \imag g\, A_\mu^c \colT{c} \,,
\end{align} 
with the Clifford algebra with
hermitian $\gamma$ matrices 
\begin{align}
	\{\gamma^\mu,\gamma^\nu\} = 2\delta^{\mu\nu}\,.
\end{align}
In general, our notation follows 
\cite{Mitter:2014wpa,Braun:2014ata,Cyrol:2016tym} of the fQCD
collaboration \cite{fQCD:2016-10}.

\subsection{FRG and dynamical hadronisation}
\label{sec:WetterichEq}

\begin{figure}[t]
	\includegraphics{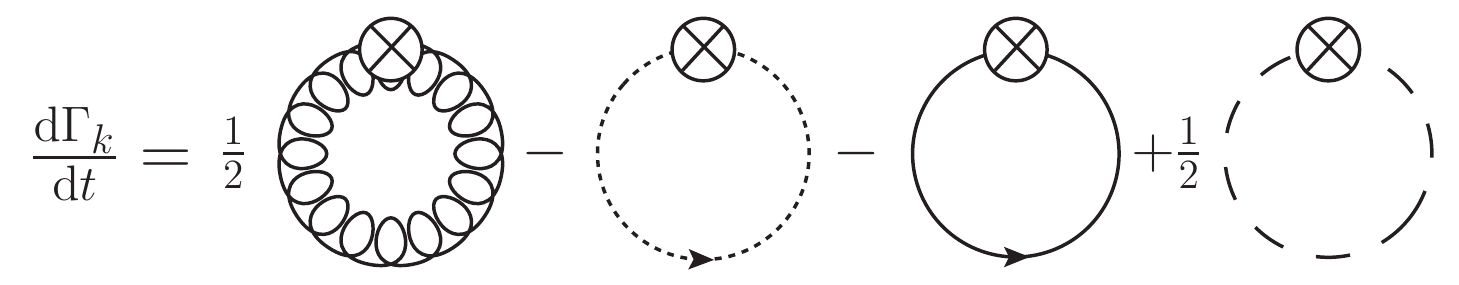}
	\caption{
		Graphical representation of the Wetterich equation.
		Lines represent the full scale-, momentum- and field-dependent 
propagators $G_k[\Phi](p)\,$, see \Fig{fig:truncation} for the line coding.
		Circled crosses represent the infrared regulator insertions 
$\partial_t R_k\,$, see \eq{eq:flow} and \eq{eq:Gn}.
	}
  \label{fig:floweq}
\end{figure}

The FRG is a non-perturbative functional method that allows to
consistently integrate quantum fluctuations in momentum shells in the
Wilsonian spirit. It relies on introducing an infrared
renormalisation group (RG)
scale that allows to interpolate between the bare action at large
RG scales and the full quantum effective action in the limit of a
vanishing RG scale, see
\cite{Litim:1998nf,Berges:2000ew,Pawlowski:2005xe,Gies:2006wv,Schaefer:2006sr,Rosten:2010vm,Braun:2011pp,vonSmekal:2012vx}
for QCD-related reviews. On a technical level this is achieved by
introducing infrared regulators $R_k$ in the dispersions of the
fields, that act like momentum-dependent mass terms and suppress
fluctuations below their effective cutoff scale. This leads to a
generating functional based on a scale-dependent classical action,
\begin{align}
	\label{eq:S_k}
	Z_k[J]=\!\int d\Phi\,e^{-S_k+\int J_\Phi\Phi}\,,\  
	S_k[\Phi] = S_{\text{\tiny{QCD}}}[\Phi]+\Delta S_k[\Phi]\,,
\end{align}  
with
\begin{align}
	\Delta S_k[\Phi]=&\,\int_x\,
		\012  A_\mu^a\, R_{k,\mu\nu}^{ab} \, A_\nu^b  
		+ \int_x\,\bar c^a\, R^{ab}_k\, c^b
	\eqnewline
	&\, + \int_x\,\bar q\, R_k^q\, q+\012 \int_x\,\phi\, R_k^\phi\, \phi\,.
	\label{eq:dSk}
\end{align} 
The super-field
\begin{align}
	\Phi=(A_\mu,c,\bar c,q,\bar q,\phi)\,,
\end{align}
includes the fundamental gluon ($A_\mu$), ghost ($\bar c, c$) and
quark ($\bar q, q$) fields as well as auxiliary hadronic degrees of
freedom ($\phi$). In the present work, the auxiliary hadronic field
represents the sigma-meson and the pions
\begin{align}
	\phi=(\sigma,\vec \pi)\,,
\end{align}
respectively. The regulator functions $R_k$
suppress the corresponding fluctuations below momentum
scales $p^2\approx k^2$ and vanish in the ultraviolet for momenta
$p^2\gg k^2\,$. See \App{app:regulator} for details on the regulators
used in the present work.

The evolution of the effective average action $\Gamma_k\,$, the
scale-dependent analogue of the effective action $\Gamma\,$, is
described by the Wetterich equation \cite{Wetterich:1992yh},
\begin{align}
	\label{eq:flow}
	\left(\partial_t +\partial_t \phi_{k}[\Phi]\0{\delta}{\delta 
	\phi}\right)\Gamma_k[\Phi] =\012 \Tr \, G_k[\Phi]\, \partial_t R_k\,, 
\end{align}
with 
\begin{align}
	\label{eq:Gn}
	G_k[\Phi]=\0{1}{\Gamma^{(2)}[\Phi]+R_k}\,,\quad \Gamma^{(n)}_{\varphi_{1}\cdots \varphi_{n}}[\Phi]
		=\0{\delta^n \Gamma_k[\Phi]}{\delta \varphi_{n}\cdots \delta \varphi_{1}}\,.
\end{align}
Here, $\varphi_i$ represents a component of the superfield $\Phi\,$, \eg
$\varphi_1=A_\mu\,$.  For a graphical representation of the flow 
equation \eq{eq:flow} see \Fig{fig:floweq}. In \eq{eq:flow} we have 
introduced $t=\log (k/\Lambda)$ which denotes the RG time. The
normalisation scale $\Lambda$ is chosen as the UV scale
$\Lambda=\Lambda_{\tiny \text{UV}}\,$. The two-point function
$G_k[\Phi]$ is the full momentum- and field-dependent
propagator in the presence of the infrared regulator $R_k\,$.

The hadronic auxiliary fields, $\phi=(\sigma,\vec\pi)\,$, are genuine,
independent fields in the effective action $\Gamma_k[\Phi]\,$, since the
latter is related to the Legendre transform w.r.t. 
$J_\Phi$. Technically, they are introduced via their flows
$\partial_t \phi_{k}[\Phi]$ as an efficient description of resonant
channels in multi-quark interactions via dynamical hadronisation
\cite{Gies:2001nw,Pawlowski:2005xe,
  Floerchinger:2009uf,Mitter:2014wpa,Braun:2014ata}. This procedure
naturally avoids the potential double-counting problems known from
low-energy effective theories. It is the source of the additional term in
the left hand side of the flow equation \eq{eq:flow}.
This additional term ensures that in each RG step the quantum corrections
in the given channel are correctly rewritten as
the exchange of the auxiliary degrees of freedom $\phi\,$.

\begin{figure}[t]
  \centering 
    \includegraphics{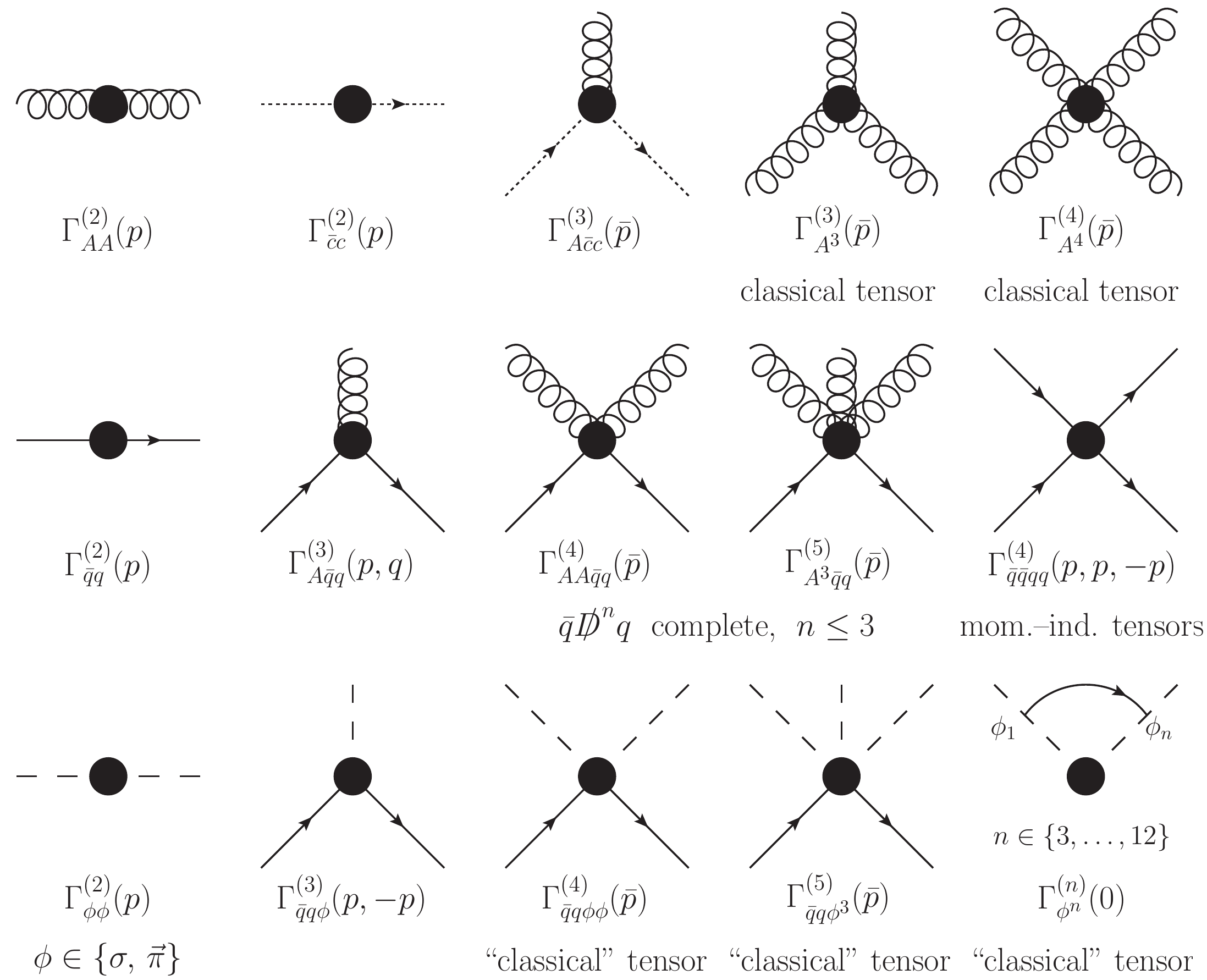}
    \caption{Vertex expansion of the effective action.
	    Wiggly lines represent gluons, dotted
      lines ghosts, solid lines quarks and dashed lines represent
      mesons introduced via dynamical hadronisation to capture
      resonant structures in four-Fermi interactions. The effective
      action is expanded about the expectation value of the scalar
      meson field, which acquires a non-vanishing value in the
      chirally broken phase. The symmetric momentum configuration 
      is denoted by $\bar{p}\,$.
      \myhfill
    }\label{fig:truncation}
\end{figure}

We emphasise that the dynamical hadronisation procedure results in a
redefinition of $1$-particle irreducibility, which is typical for the
introduction of dynamical effective degrees of freedom. In the present
case, a former $1$PI four-Fermi interaction is transformed into the
$1$-particle reducible exchange of mesons. Amongst other advantages,
this allows us to efficiently include the effects of the
multi-scatterings of the resonant channels via the inclusion of the
corresponding higher interactions, \cf the bottom line of
\Fig{fig:truncation}. Finally, the auxiliary fields can be removed
again via their equations of motion
$\delta\Gamma_k/\delta\phi[\phi=\phi_{\text{\tiny{EoM}}}]=0$:
\begin{align}
	\label{eq:GeffGQCD}
	\Gamma_{{\rm QCD},k}[A,c,\bar c,q,\bar q]=\left.\Gamma_k[\Phi]\right|_{\phi=\phi^{\ }_{\text{\tiny{EoM}}}}\,, 
\end{align}
resulting in the standard effective action in terms of the fundamental
QCD degrees of freedom at vanishing cutoff scale. Since the
fields $\phi$ are related to bilinears of the quarks, the mesonic
cutoff term introduces an ultraviolet vanishing form factor to the
scalar--pseudo-scalar channel of the four-Fermi interaction for
$k > 0$. This term acts as an infrared regularisation for resonant
interaction, which leaves the renormalisability of the theory
unspoilt.

\subsection{Vertex expansion and truncation}
\label{sec:vertexexpansion}

We approximate the flow equation \eq{eq:flow} within a vertex
expansion scheme, \ie an expansion of the effective action in terms
of $1$PI correlation 
functions. Functional derivatives of the equation
with respect to the fields $\varphi_1,\dots,\varphi_n$ lead to functional 
equations for $1$PI $n$-point functions, parametrised by
\begin{align}
	\label{eq:proper_vertices}
	\Gamma^{(n)}_{\varphi_1\dots\varphi_n} &= \sum\limits_{i} \lambda^{(i)}_{\varphi_1\dots\varphi_n}{\cal T}^{(i)}_{\varphi_1\dots\varphi_n}\,.
\end{align}
Here, the $\lambda^{(i)}$ represent dressing functions and the
${\cal T}^{(i)}$ represent a tensor basis for the corresponding proper
vertex.  For vertices that appear in the classical action, the
classical tensor structure corresponds to the index $i=1$. The
above parametrisation of the proper vertices differs from the
RG-invariant parametrisation used in
\cite{Mitter:2014wpa} by factors of the scalar
propagator dressing functions.

\begin{figure}[t]
	\includegraphics{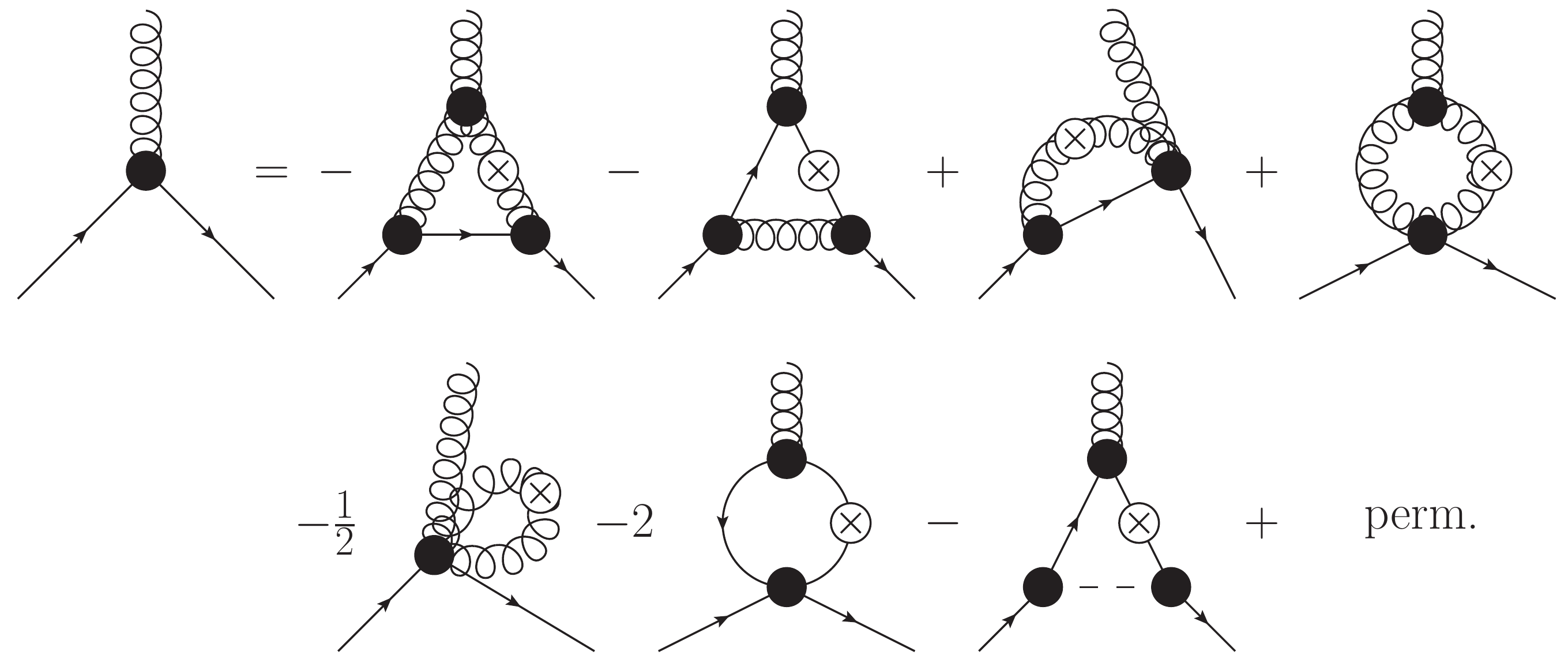}
	\caption{Diagrammatic representation of our approximation of
          the flow equation for the quark-gluon vertex.  Permutations
          include \mbox{(anti-)symmetric} permutations of external
          legs as well as permutations of regulator insertions.}
	\label{fig:gql_floweq}
\end{figure}

The flow equation of any vertex
$\Gamma^{(n)}_{\varphi_1\dots\varphi_n}$ depends on higher correlation
functions up to $(n+2)$-point functions.  In order to render the
system numerically tractable, the infinite tower of coupled equations
for the correlation functions has to be truncated at some finite
order.  The truncation used in this work builds on the truncations
used in the previous works in quenched two-flavour QCD
\cite{Mitter:2014wpa} and Yang-Mills theory \cite{Cyrol:2016tym}.
Therefore, we discuss only the general expansion scheme and those
constituents of our truncation, \cf \Fig{fig:truncation}, that have
not been included in these works.  In particular, the
latter are given by the two-quark-two-gluon, the
two-quark-three-gluon, the two-quark-two-meson as well as the
two-quark-three-meson vertices. Their representation in terms of
tensors is presented in \App{app:quarkvertices}.  Due to the large
size of the truncation, we refrain from giving explicit diagrammatic
representations for the flow equations of the involved propagators and
vertex functions. As a representative example, we show the flow
equation for the quark-gluon vertex in \Fig{fig:gql_floweq} and refer
the reader to \cite{Mitter:2014wpa,Cyrol:2016tym} for further flow
equations. The computer algebraic tool DoFun \cite{Huber:2011qr}
allows to obtain the corresponding diagrams with little effort.

\begin{table}[b]
  \centering
\begin{tabular}{c||ccc} 
	back-coupled into & classical & leading & sub-leading\\\hline
	classical & $\checkmark$ & $\checkmark$ & $\checkmark$\\
	leading & $\checkmark$ & $\checkmark$ & $\times$\\
	sub-leading & $\checkmark$ & $\times$ & $\times$
\end{tabular}
\caption{
	Back-coupling of tensor classes. For the classification of the tensors and
	a discussion of the deviations from this scheme see \App{app:truncationscheme}.
	}
\label{tab:backcoupling}
\end{table}

In order to make the systematics of the vertex expansion apparent, we sort the
constituents of our truncation into three groups,
\begin{itemize}
	\item classical tensors,
	\item leading non-classical tensors,
	\item sub-leading non-classical tensors,
\end{itemize}
where the assignment of tensors to the three groups is specified in
\App{app:truncationscheme}. 

In the present implementation of this scheme we have assumed that
leading non-classical tensors which are plugged into the equations of
sub-leading non-classical tensors have only a sub-sub-leading overall
effect. Analogously, we have also assumed the same for sub-leading non-classical
tensors, if they are plugged into the equations of leading
non-classical tensors.
The above assumptions have been 
verified in many
cases.
In particular, the sub-leading non-classical tensors 
have been found to yield only sub-sub-leading corrections 
to the flows of the leading non-classical tensors of 
the quark-gluon vertex, cf. \cite{Mitter:2014wpa}.
Apart from a few exceptions discussed in
\App{app:truncationscheme}, we finally take into account only contributions up to
the sub-leading level, which is illustrated in
\Tab{tab:backcoupling}.  The resulting truncation consists of
the largest set of correlation functions that has so far been solved 
with functional methods. Nevertheless, a careful assessment of
truncation artefacts is essential. We discuss the convergence, the
classification of tensors, and the approximation of the momentum
dependence of the different dressing functions in detail in
\Sec{sec:results}, \App{app:truncationscheme} and
\App{app:quarkvertices}. Furthermore, the source and assessment of the
leading truncation error is discussed in the next section as well as
in \Sec{sec:AppCon}.

\subsection{STI of the quark-gluon vertex} 
\label{sec:STI}

In the resummation scheme defined by the FRG, spontaneous chiral
symmetry breaking is triggered by the dynamical creation of a
four-Fermi interaction from box-diagrams with two exchanged gluons,
see \eg \cite{Braun:2011pp}.  The presence or absence of
spontaneous chiral symmetry breaking is therefore most sensitive to the strength
of the momentum-dependent quark-gluon vertex interaction
\cite{Mitter:2014wpa}. Consequently, even small quantitative errors
can have devastating effects on the qualitative and quantitative
behaviour of the mechanism of dynamical chiral symmetry breaking. 
In order to minimise this sensitivity, we use the Slavnov-Taylor identity 
to constrain the quark-gluon vertex. At the symmetric momentum
configuration, the STI leads to the identity, see
\eg \cite{Davydychev:2000rt,Aguilar:2010cn,Rojas:2013tza,Aguilar:2014lha,Aguilar:2016lbe},
\begin{align}
	&\lambda^{(9)}_{\bar qq A}(\psym) = \0{Z_q(\psym)}{Z_c(\psym)}\left[\lambda^{(1)}_{cq Q_q}(\psym)-
	\frac{3}{2}\, \psym^2\, \lambda^{(4)}_{cq Q_q}(\psym)\right]\,,
	\label{eq:STI}
\end{align}
see also \App{app:quarkghostscatteringkernel} for more details. Here,
$\lambda^{(9)}_{\bar qq A}(\psym)$ is the dressing function of the
longitudinally projected classical tensor structure of the quark-gluon
vertex, \cf \App{app:quarkvertices}. {The generalised BRST-vertices
$\lambda^{(1)}_{cq Q_q}(\psym),\,\lambda^{(4)}_{cq Q_q}(\psym)\,$, see
\eq{eq:qcQq}, are field derivatives of the quantum BRST variation of
the quark, see \eq{eq:qcQq} and \App{app:quarkghostscatteringkernel} for 
more details.
 
Note also that the BRST-vertices are fully dressed, and can be
obtained via their respective flow equation,
\eq{eq:flowBRST}. Accordingly, the STI \eq{eq:STI} only contains fully
dressed quantities that can be derived from the initial effective
action via their flows. This leaves us with a consistent set of flows
and STIs , where each relation by itself has the correct cutoff and
RG scaling. In particular the important self-consistency of the
renormalisation procedure is guaranteed.

Finally, the right hand side of \eq{eq:STI} implicitly introduces the
propagator coupling $\alpha_s$
\cite{vonSmekal:1997ohs,vonSmekal:1997ern,Hauck:1998fz},
\begin{align}
	\label{eq:propcoupling}
	\alpha_s(\psym)= \0{1}{4\pi} \,\0{g^2 }{Z_A(\psym) Z_c^2(\psym)}\,,
\end{align}
where $Z_A$, $Z_q$ as well as $Z_c$ are the wave function renormalisations
of the gluon, quark and ghost propagators, 
\cf \eq{eq:propagators}. The renormalised coupling $g$ is
defined at the momentum $p_{\text{\tiny{ren}}}$ with
$Z_A(p_{\text{\tiny{ren}}}) Z_c^2(p_{\text{\tiny{ren}}})=1$ with
$g^2 = 4 \pi\alpha_s(p_{\text{\tiny{ren}}})\,$, setting the
renormalisation scale. Typically this scale is close to the initial
cutoff scale $k=\Lambda\,$. With \eq{eq:propcoupling} we can rewrite the
ratio of wave function renormalisations in \eq{eq:STI} as
\begin{align}
	\label{eq:normalSTI}
	\0{Z_q(\psym)}{Z_c(\psym)} = \0{\sqrt{4 \pi\alpha_s(\psym)}}{g} Z_q(\psym)Z^{1/2}_A(\psym)\,.
\end{align}
This makes the RG scaling of the quark-gluon vertex apparent at the
expense that it seemingly does depend on the renormalised coupling
$g\,$, which is not a fully dressed quantity.

In the literature, the STI \eq{eq:STI} is often used to constrain 
the leading dressing function $\lambda^{(1)}_{\bar qq A}$ of the transversely
projected quark-gluon vertex via the identification
\begin{align}
	\label{eq:STI_condition}
	\lambda^{(1)}_{\bar qq A}(p,q)\equiv\lambda^{(9)}_{\bar qq A}(p,q)\ ,
\end{align}
where $\lambda^{(1)}_{\bar qq A}$ and $\lambda^{(9)}_{\bar qq A}$ are the dressing function of the transversely 
and longitudinally projected classical tensor structure in our basis, \cf \App{app:quarkvertices}.
However, there are some crucial assumptions being made in this identification, which
is obviously admissible at the classical level. As discussed in more detail in the appendix,
the assumption of generalised regularity, in the sense that the longitudinal and transverse
projection of the generators for the full basis \eq{eq:barqAq_tensors} are not independent, results in the
more general relation
\begin{align}
	\label{eq:qgl_long_trans_relation}
  \lambda&^{(9)}_{\bar q q A}(p,q)=\\
         &\left[\lambda^{(1)}_{\bar q q A}(p,q)+\left(\frac{
           \lambda^{(7)}_{\bar q q A}(p,q)}{2}-\lambda^{(5)}_{\bar q q A}(p,q)\right)
           \left(p^2-q^2\right)\right]\ .\nonumber
\end{align}
However, it was found already in \cite{Mitter:2014wpa} that the relation
\begin{align}
	\label{eq:qgl_trans_relation}
	\frac{\lambda^{(7)}_{\bar q q A}(p,q)}{2}=\lambda^{(5)}_{\bar q q A}(p,q)\,,
\end{align}
is fulfilled to very high precision at momenta larger than
\SI{1}{\GeV}.  Consequently, the assumption of generalised regularity
as well as the equality \eq{eq:qgl_trans_relation} are needed in order
to guarantee the validity of identification \eq{eq:STI_condition} and
lie therefore at the heart of any application of the Slavnov-Taylor
identity to the transversely projected classical tensor structure of
the quark-gluon vertex.  In particular, the results presented in
\cite{Mitter:2014wpa} show that \eq{eq:qgl_trans_relation} is clearly
violated at non-perturbative momenta, therefore invalidating the usage
of the STI to constrain $\lambda^{(1)}_{\bar q q A}$ via
\eq{eq:STI_condition}.

\begin{figure*}[t]
  \centering \subfloat[ Gluon propagator dressing function $1/Z_A(p)$
  in comparison to lattice results \cite{Sternbeck:2012qs}.  The
  discrepancy at large momenta stems from lattice discretisation
  artefacts.\myhfill] {
		\includegraphics{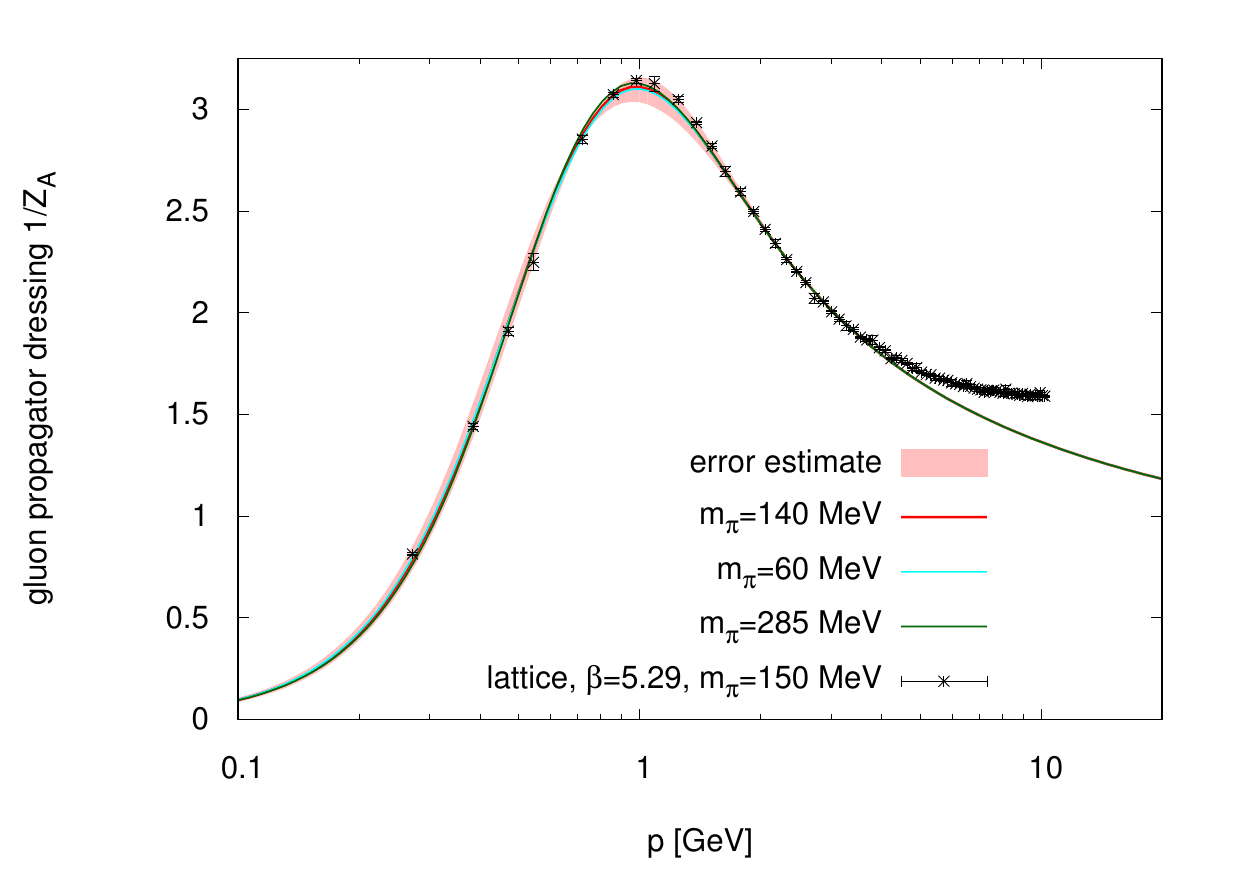}
		\label{fig:qcd2_glueprop_mpi}
              } \hfill \subfloat[ Quark propagator dressing functions,
              $M_q(p)$ (lower line, \si{\GeV}) and $1/Z_q(p)$ (upper line),
              in comparison to lattice results
              \cite{Oliveira:2016muq, *Sternbeck:PC17}.  We do not show lattice results
              for the quark wave function $Z_q$ since the currently
              available two-flavour results are still plagued by
              considerable systematic errors.\myhfill] {
		\includegraphics{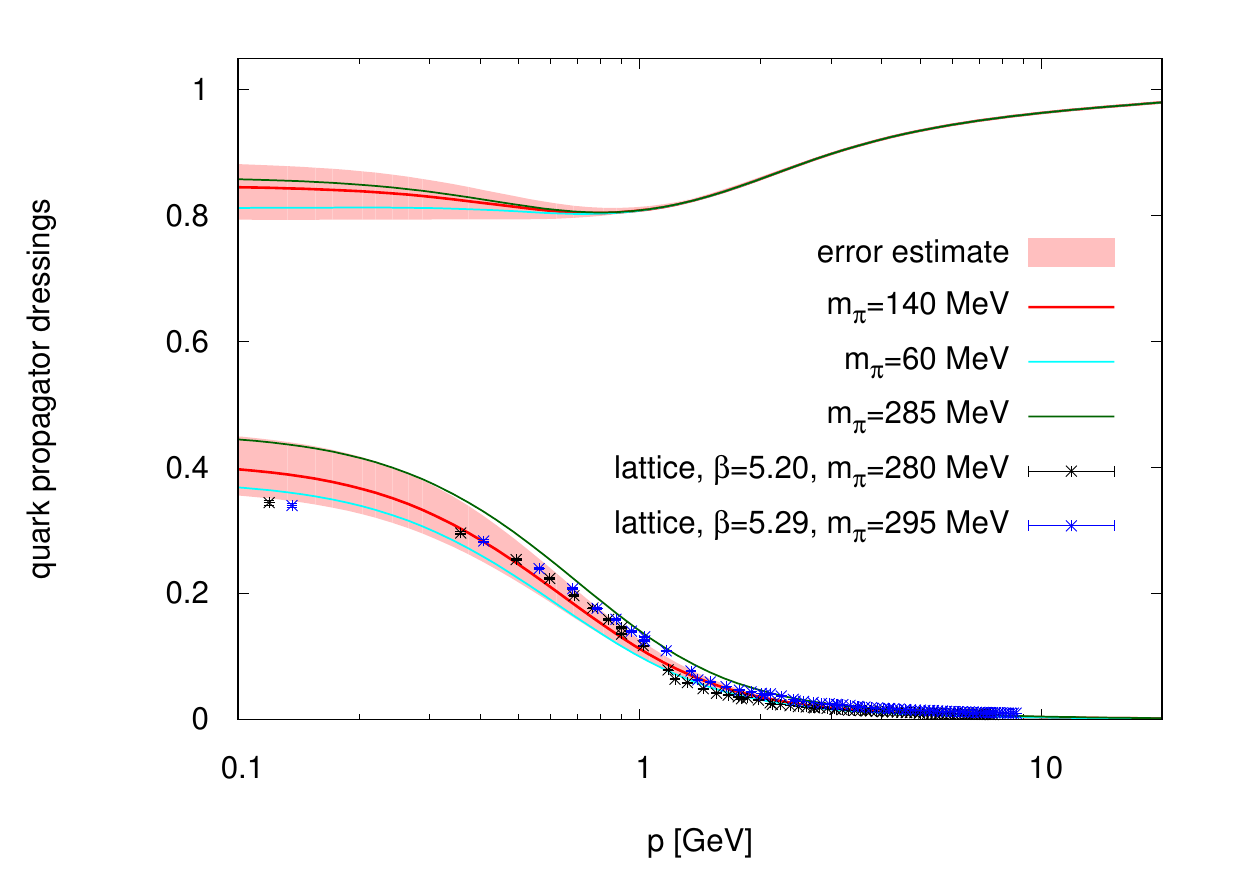}
		\label{fig:quark_prop_mpi}
	}
	\caption{Two-flavour unquenched gluon and quark propagators
          for different pion masses in comparison to lattice results.}
	\label{fig:propagator_mpi}
\end{figure*}

To understand the implications of the assumption of generalised
regularity we look at the quark-gluon vertex in the limit of vanishing
gluon momentum. Due to the structure of the longitudinal and
transverse projection, a violation of \eq{eq:STI_condition} at small
gluon momentum,
\begin{align}
	\label{eq:irregularity}
  \lim\limits_{q\rightarrow -p}\lambda^{(1)}_{\bar qq A}(p,q){\neq}\lim
  \limits_{q\rightarrow -p}\lambda^{(9)}_{\bar qq A}(p,q)\,,
\end{align}
implies an irregularity of the full quark-gluon vertex at vanishing gluon
momentum.
Note in this context that the dynamical creation of the gluon
mass gap, and hence that of confinement
\cite{Braun:2007bx,Fister:2013bh}, indeed requires
irregularities in the vertices, see \cite{Cyrol:2016tym} for a
detailed discussion. Even though this does not necessarily imply an
irregularity of the quark-gluon vertex, the gapping scale for the
gluon propagator provides another estimate for the scale below which
the identification \eq{eq:STI_condition} is no longer enforced by 
the STI.

From these regularity arguments and the numerical finding
\eq{eq:qgl_trans_relation}, we conclude that there exists a
scale,
\begin{align}
	\label{eq:scale_estimate}
	\LambdaSTI = \mathcal{O}(\SI{1}{\GeV})\,,
\end{align}
below which \eq{eq:STI_condition} cannot be safely applied any
more. To obtain a better estimate of the STI scale
$\LambdaSTI$, we consider the transverse running
couplings extracted from the different gluonic vertices,
\begin{align}
	\label{eq:runcoup}
	\alpha_{\bar c c A}(\psym) &= \0{1}{4 \pi}\,\frac{\left(\lambda^{(1)}_{\bar cc A}(\psym)\right)^2}
	{ Z_A(\psym)\,Z_c^2(\psym)}\,,\\\nonumber
	\alpha_{A^3}(\psym) &= \0{1}{4 \pi} \,\frac{\left(\lambda^{(1)}_{A^3}(\psym)\right)^2}
	{Z_A^3(\psym)}\,\\\nonumber
	\alpha_{A^4}(\psym) &= \0{1}{4 \pi}\, \frac{\lambda^{(1)}_{A^4}(\psym)}{
	Z_A^2(\psym)}\,.
\end{align} 
Here, $\lambda^{(1)}_{\varphi_1\dots \varphi_n}$ are the dressing
functions of the classical tensors of the vertices
$\Gamma^{(n)}_{\varphi_1\dots \varphi_n}$ as defined in
\eq{eq:proper_vertices}. As in the case of the quark-gluon vertex, the
corresponding STIs for the gluonic vertices constrain only their
longitudinal projection. Assuming negligible non-classical tensor
structures, the transverse running couplings \eq{eq:runcoup} show
therefore degeneracy as long as a condition analogous to
\eq{eq:STI_condition} is valid for each of the gluonic
vertices. Therefore, the scale at which the degeneracy in the
transverse gluonic couplings is lifted yields an approximation for
$\LambdaSTI\,$.  Based on our results for the transverse gluonic running
couplings, see \Fig{fig:running_coupling}, we identify the scale where
degeneracy is lost as
\begin{align}
	\label{eq:STI-scale}
	\LambdaSTI \lesssim 3-5\, {\rm GeV}\,.
\end{align}
In particular, the degeneracy of the gluonic running couplings is
violated by more than \SI{5}{\percent} below \SI{3}{\GeV}, and the
STIs cannot be reliably used to constrain transversely projected
couplings below this scale. On the other hand, we observe
near-degeneracy of the gluonic couplings above \SI{5}{\GeV}.

With the identification of the scale $\LambdaSTI$ we are now in a
position to apply the STI \eq{eq:STI} to constrain the quark-gluon
vertex, for which a transverse running coupling can be defined via
\begin{align}
	\label{eq:runcoupqgv}
  \alpha_{\bar q q A}(\psym) = \0{1}{4 \pi}\,\frac{
  \left(\lambda^{(1)}_{\bar qqA}(\psym)\right)^2}{
  Z_A(\psym)\,Z_q^2(\psym)}\,.
\end{align}
We use \eq{eq:STI} together with \eq{eq:STI_condition} to calculate
specific momentum configurations of the dressing function
$\lambda^{(1)}_{\bar qq A}$ of the transversely projected classical
tensor structure ${\cal T}^{(1)}_{\bar qq A}(\psym)$ of the
quark-gluon vertex. Due to the presence of the RG scale,
$\LambdaSTI$ plays a two-fold r\^ole.
For RG scales $k>\LambdaSTI\,$, we use the STI to constrain the full
range of symmetric momenta $\psym\in[0,\infty)\,$. All other momentum
configurations of $\lambda^{(1)}_{\bar qq A}(p,q)$ are calculated as
relative offset to this line of symmetric momentum configurations,
whereas the non-classical dressings $\lambda^{(i)}_{\bar qq A}(p,q)$
for $i>1$ are always calculated from the vertex equation.
For RG scales $k\lesssim \LambdaSTI\,$, on the other hand, we use the
STI to constrain only the restricted range of symmetric momenta
$\psym \in[\LambdaSTI,\infty)$ of $\lambda^{(1)}_{\bar qq
  A}(p,q)\,$. Again all other momentum configurations are calculated as
relative offsets, whereas all non-classical dressings are fully
calculated from the vertex equation.  The dependence of our results on
varying the exact transition scale $\LambdaSTI$ for $k$ and $\bar p$
within a range of $3-7$ \si{\GeV} beyond the estimate \eq{eq:STI-scale}
leaves us with an estimate of our truncation error.  This error is
indicated by the bands in our results. The solid lines in our results
correspond to the upper value of \SI{5}{\GeV} for the transition scale in
\eq{eq:STI-scale}.

\subsection{Renormalisation and numerical computation}

\begin{figure*}[t]
	\centering \subfloat[Gluon propagator $1/(p^2 Z_A(p))\,$.]{
		\includegraphics{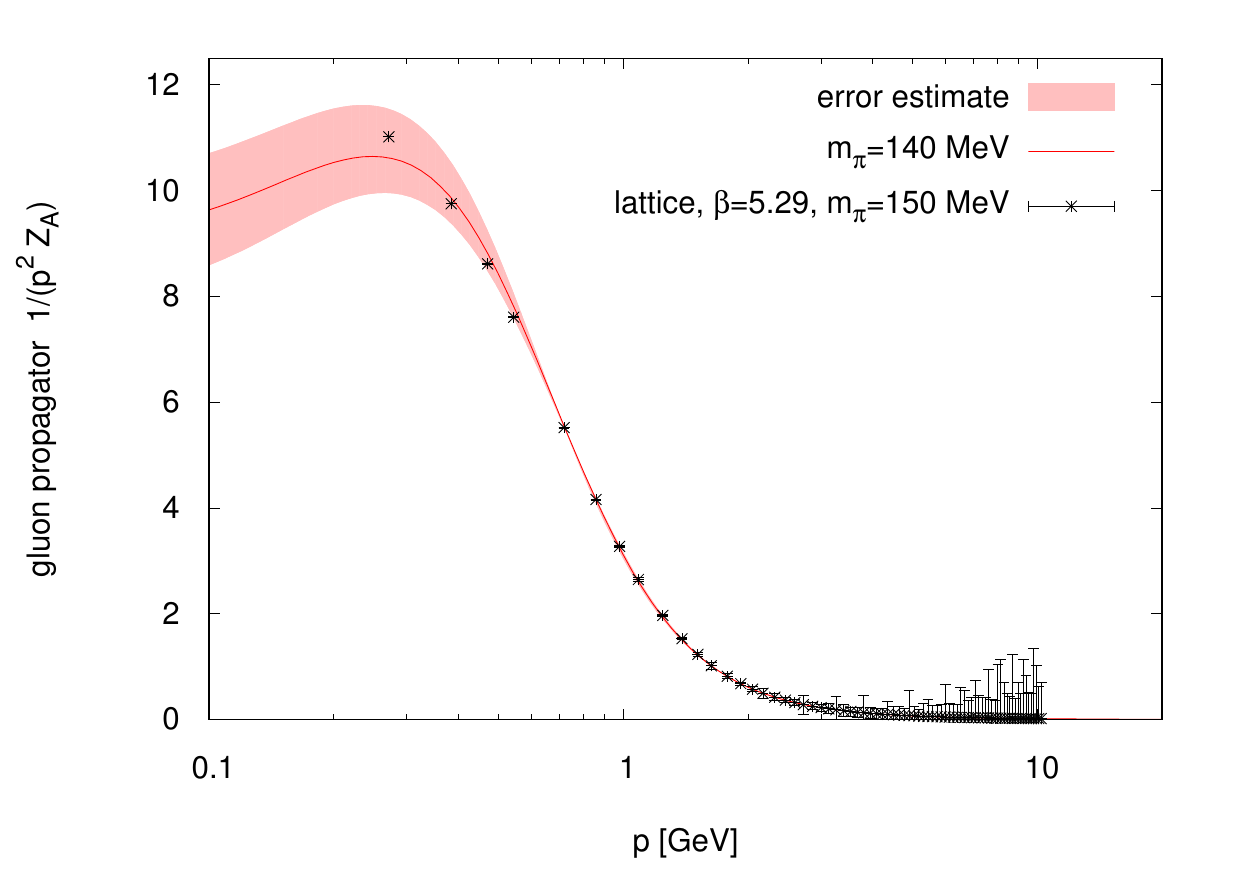}}\hfill
	\subfloat[Ghost propagator dressing function $1/Z_c(p)\,$.]{
		\includegraphics{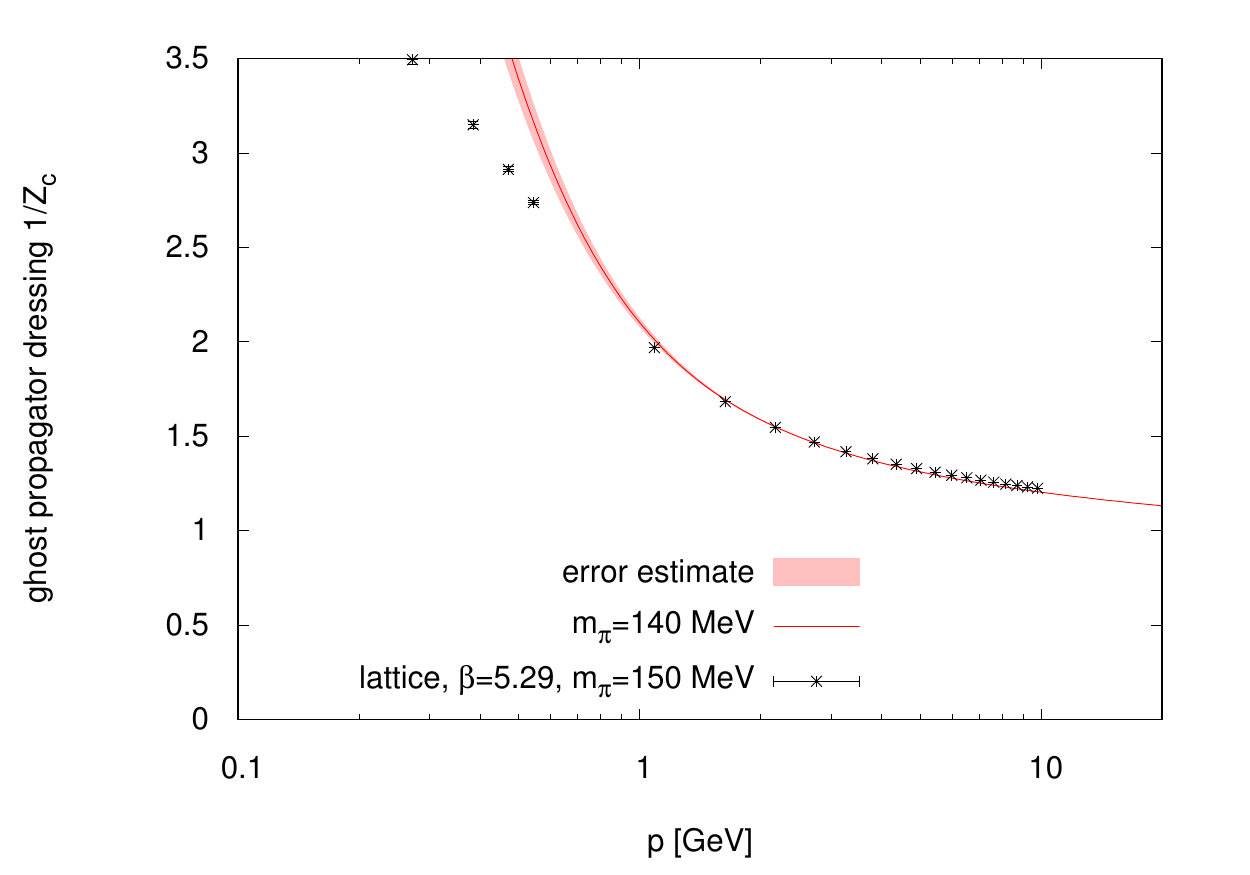}
		\label{fig:ghostprop}}
              \caption{Two-flavour gluon and ghost propagators in
                comparison to lattice data \cite{Sternbeck:2012qs}.}
	\label{fig:gluepropagators}
\end{figure*}

The solution of the flow equation starts from an initial action
$\Gamma_\Lambda$ at some large momentum cutoff $\Lambda\,$. For
$\Lambda\to \infty$ the initial action turns into the classical bare
action $\Gamma_{\Lambda\rightarrow\infty}=S$ in the presence of a
momentum cutoff. The latter implies that the bare action includes besides
the classical terms also a cutoff-induced mass term for
the gluon, \cf \cite{Cyrol:2016tym} and references therein.  

In order to minimise cutoff artefacts while keeping the
numerical effort at a manageable level, we calculate our initial
action at $\Lambda=\SI{20}{\GeV}$ in a simpler truncation. Starting 
from the
renormalised classical action at $\Lambda=\SI{100}{\GeV}$, we determine
$\Gamma_{\Lambda=\SI{20}{\GeV}}$ in a truncation that takes only the
propagators and classical vertex tensor structures into
account. Furthermore, we choose the renormalisation constants at the
cutoff scale $\Lambda=\SI{100}{\GeV}$ such that the resulting vertices
$\Gamma^{(n)}_{k\rightarrow 0}(\psym)$ fulfil the Slavnov-Taylor
identities at the symmetric momentum configuration at 
$\psym=\SI{10}{\GeV}$. 

Having calculated the initial action
$\Gamma_{\Lambda=\SI{20}{\GeV}}$ in this fashion, we integrate the full
set of equations down to the infrared cutoff of $\Lambda_{\rm 
IR}=\SI{70}{\MeV}$, where the bare quark mass at \SI{20}{\GeV} is chosen 
such that the
desired pion mass is achieved, \cf \cite{Mitter:2014wpa}.  The gluon
mass parameter, which is present at $k>0$ due to the violation of BRST
symmetry by the regularisation, is uniquely determined by the scaling
condition in the glue sector, see \cite{Cyrol:2016tym} for further
details.  For numerical convenience, the coupled matter-glue system is
solved in an iterative procedure. 
Starting point is a solution of the full system with massless quarks.
The scale dependent gluonic correlators of this solution are then fed back
as input into the matter system.
The resulting matter propagators and vertices are in turn
used as input for the glue system. This process is repeated until 
convergence is obtained.

The above procedure has been described in physical units \si{\GeV}.
In practical first-principle calculations one chooses a value for the strong
running coupling at the renormalisation scale and translates into \si{\GeV} only afterwards.
In order to facilitate the comparison with the lattice, we use the location of the
bump in the gluon propagator dressing to set our scale to lattice units
which are given in \si{\GeV}.

The general computational framework is the one presented in
\cite{Mitter:2014wpa,Cyrol:2016tym} and the reader is referred to these works for
additional details.  The algebraic flow equations are derived using
\textit{DoFun}~\cite{Huber:2011qr} and subsequently traced with
\textit{FormTracer}~\cite{Cyrol:2016zqb}, a \textit{Mathematica} package 
that uses FORM \cite{Vermaseren:2000nd,Kuipers:2012rf,Kuipers:2013pba}. The 
output is exported as optimised C code into the \textit{frgsolver}, a 
flexible, object-orientated, parallel C++ framework for the solution of 
flow equations, developed within the fQCD collaboration 
\cite{fQCD:2016-10}.

\section{Results}
\label{sec:results}

\begin{figure*}[t]
  \centering \subfloat[Dressing functions in the soft gluon limit,
  $\lambda^{(i)}_{\bar q q A}(p,-p)\,$.  The classical tensor
  structure $\lambda^{(1)}_{\bar q q A}(p,-p)$ is compared to lattice
  data \cite{Oliveira:2016muq, *Sternbeck:PC17} and normalised to match our results at
  \SI{1}{\GeV}.
	\label{fig:quarkgluon}
	\myhfill]{\includegraphics{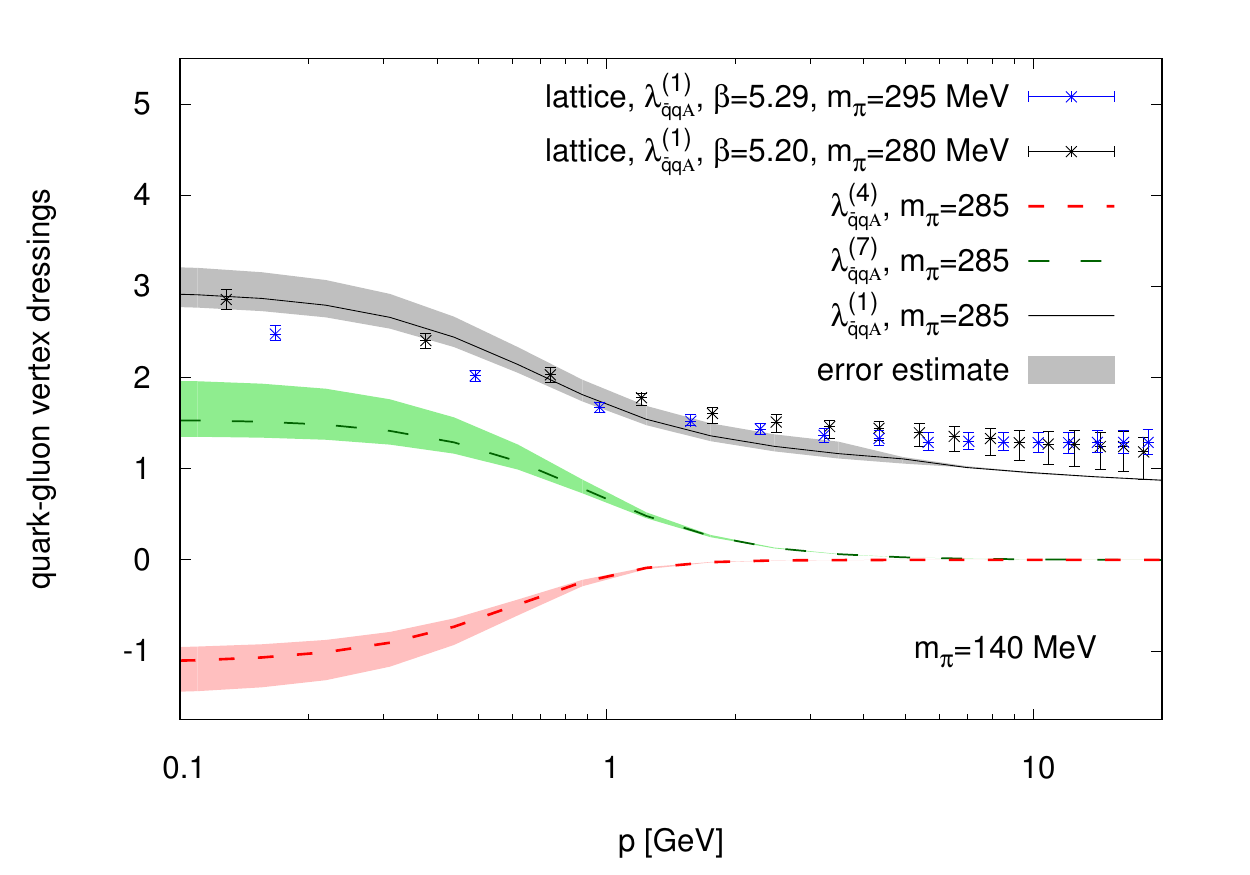}} \hfill
        \subfloat[Dependence on relative and angular momentum
        variables for fixed momentum scales
        $\bar{p}=\sqrt{(p_1^2+p_2^2+p_3^2)/3}\,$. Lines correspond to
        the symmetric point momentum configuration
        $\|p_1\|=\|p_2\|=\|p_3\|$.\myhfill]{
		\includegraphics{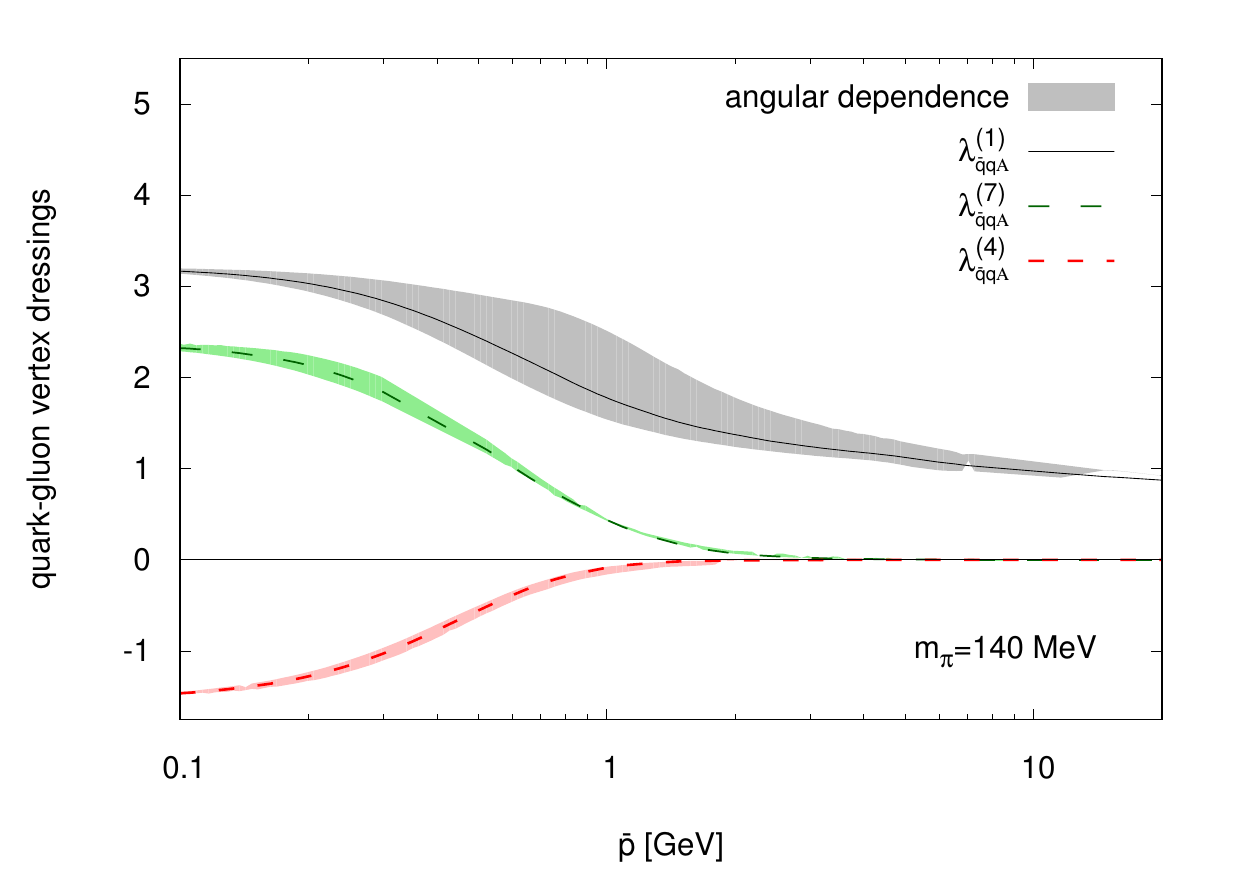}
		\label{fig:quarkgluon_angdep}
	}
	\caption{Two-flavour quark-gluon vertex dressing functions 
		$\lambda^{(i)}_{\bar q q A}\,$.}
	\label{fig:main_result1}
\end{figure*}

As a main result of our investigation, we present in 
\Figs{fig:propagator_mpi}{fig:gluepropagators} the unquenched gluon,
quark and ghost propagators. The inverse propagators are parametrised by
\begin{align}
	\label{eq:propagators}
	&\left[\Gamma_{AA}^{(2)}\right]^{\mu\nu}(p) = 
		Z_A(p)\, p^2\, \left(\delta^{\mu\nu}-\frac{p^\mu p^\nu}{p^2}\right)\,,\eqnewline
	&\left[\Gamma_{\bar qq}^{(2)}\right](p) =
	 Z_q(p)\, \left(\imag\slashed{p} + M_q(p)\right)\ ,\eqnewline
	&\left[\Gamma_{\bar cc}^{(2)}\right](p) = Z_c(p)\, p^2\,,
\end{align}
with suppressed colour and flavour indices for notational
simplicity. Note that the dressing functions introduced in
\eq{eq:propagators} are the inverse of the dressing functions $Z$ and
$G$ often used in the DSE literature to parametrise the gluon and
ghost propagators, whereas $Z_q$ corresponds to the $A$ function,
often used to parametrise the quark propagators, see
e.g.\ \cite{Alkofer:2000wg,Eichmann:2016yit}. Our results have been
obtained with $N_f=2$ quark flavours at different pion masses. We
indicate our best estimate of the systematic error due to truncation
effects via bands, \cf the discussion in \Sec{sec:STI} and
\Sec{sec:AppCon}.

The gluon propagator dressing function $1/Z_A(p)\,$, shown in
\Fig{fig:qcd2_glueprop_mpi}, shows unprecedented agreement with
unquenched two-flavour lattice results
\cite{Sternbeck:2012qs}. The latter result flattens out due to
lattice artefacts at large momenta, while our gluon propagator shows
a smooth transition to the expected perturbative behaviour.  
At very small momenta we find small deviations to the
lattice result, as our gluon propagator is of the scaling type, \cf
\cite{Cyrol:2016tym}, whereas the lattice results are of the
decoupling type.
Note in this context that the effects due to the non-perturbative
gauge-fixing procedure are still an open issue, see \eg
\cite{Maas:2009se,Sternbeck:2012mf,Maas:2015nva,Maas:2017csm}.
Therefore, any comparison of correlators should keep such
effects in mind.  It is also noteworthy that the gluon propagator is
insensitive to the pion mass. This insensitivity to the details
of the matter sector is a very welcome property for investigations
of the phase structure of QCD at finite temperature and
density. There we expect significant changes in the dynamics of the
matter sector, whose impact on the glue sector should be
limited by the above mechanism. Consequently, this stabilises the
current vertex expansion scheme at finite temperature and density. 
In particular, these findings strengthen the predictive power
of approaches that use lattice input for the gauge sector of the 
truncation \cite{Fischer:2011mz,Fischer:2012vc,Fischer:2014ata}.

Our result for the quark propagator is shown in
\Fig{fig:quark_prop_mpi}.  At intermediate and large momenta we find
very good agreement of the quark mass function, $M_q(p)\,$, with
corresponding lattice results \cite{Oliveira:2016muq, *Sternbeck:PC17}.  However, we
find a larger infrared value for the quark mass function as compared
to the lattice. As discussed in more detail in \Sec{sec:AppCon}, this
is most likely an artefact of the presence of a slight scale mismatch
between the matter and glue sector.  We refrain from a comparison of
the quark wave function renormalisation to the lattice since presently
no two-flavour continuum-extrapolated results with reliable systematic
errors are available. It is interesting to compare the qualitative
behaviour of $Z_q$ with other functional method calculations. We find
a slight backbending of the quark wave function renormalisation at
small momentum scales. A similar, but more pronounced, effect has also
been observed in Dyson-Schwinger studies of the quark propagator, see \eg
\cite{Fischer:2008sp,Fischer:2008wy,Aguilar:2010cn,Williams:2014iea,Aguilar:2016lbe}.
We find a decreased backbending for a smaller pion mass, see
\Fig{fig:quark_prop_mpi}.  This is the opposite effect to the one
found in
\cite{Fischer:2008sp,Fischer:2008wy,Aguilar:2010cn,Williams:2014iea,Aguilar:2016lbe}.
On the other hand, the quark mass function, $M_q(p)\,$, shows the
expected monotonic dependence on the pion mass.

Apart from the gluon propagator, we show a comparison of our results
for the ghost propagator dressing $Z_c(p)$ to the lattice results of
\cite{Sternbeck:2012qs} in \Fig{fig:gluepropagators}. Similarly to
pure Yang-Mills theory \cite{Cyrol:2016tym}, the scaling ghost
propagator agrees with the lattice decoupling solution only down to
momenta of about \SI{1}{\GeV}.  On the other hand, although we have
a scaling solution, our gluon propagator agrees remarkably well with
the decoupling lattice propagator down to comparably low momenta.

The self-consistent solution of our large truncation provides us with
a wealth of non-trivial information on vertex functions.  This
includes in particular the momentum dependence of classical and
non-classical tensor structures, many of which are calculated here for
the first time.  Here, we focus on a detailed discussion of our
results for the quark-gluon vertex as the most crucial ingredient for
quantitative accuracy in the unquenched system.  The transversely
projected quark-gluon vertex can be represented with eight basis
elements \cite{Ball:1980ay}. They include four chirally symmetric
tensors, one of them being the classical tensor, as well as four
tensors which break chiral symmetry, see \App{app:quarkvertices}.  In
line with earlier investigations
\cite{Hopfer:2013np,Williams:2014iea,Mitter:2014wpa,Williams:2015cvx}, it turns out
that only two non-classical tensor structures have to be considered as
leading non-classical tensors in the backcoupling-scheme shown in
\Tab{tab:backcoupling}, see also the detailed discussion of the
truncation scheme in \App{app:truncationscheme}.  The first, and
quantitatively most important, is the chirally symmetric tensor
structure $\mathcal{T}^{(7)}_{\bar q q A}\,$, the second is given by
the chiral symmetry-breaking tensor structure
$\mathcal{T}^{(4)}_{\bar q q A}$, see \eq{eq:barqAq_tensors} for the
considered basis.  Our results for the leading dressing functions of
the quark-gluon vertex are shown in \Fig{fig:quarkgluon} in comparison
to the lattice results for the classical tensor structure
\cite{Oliveira:2016muq, *Sternbeck:PC17}.  Within the errors we find good agreement
with the lattice results in the soft-gluon limit. Consistent with
earlier investigations \cite{Mitter:2014wpa}, we find that the
dressing of the classical tensor structure of the quark-gluon vertex
shows a sizeable angular dependence, as illustrated in
\Fig{fig:quarkgluon_angdep} and \Fig{fig:quarkgluon_3D}.  We checked
that this angular dependence is genuine and cannot be simply removed
by a re-parametrisation with propagator dressings.  Therefore, the
inclusion with the full three-dimensional momentum-dependence is
required.  This is in contrast to the gluonic vertices, where
one-dimensional momentum approximations at the symmetric point
represent already a quantitatively good approximation, see
\cite{Cyrol:2014kca,Cyrol:2016tym} and \App{app:truncationscheme}.
The chirally symmetric tensor structure
$\mathcal{T}^{(7)}_{\bar q q A}$ takes sizeable values already in the
semi-perturbative momentum region whereas
$\mathcal{T}^{(4)}_{\bar q q A}$ is of quantitative importance only in
the chirally broken phase.

\begin{figure*}[t]
  \centering \subfloat[Classical tensor structure
  $\lambda^{(1)}_{\bar q q A}$ of the quark-gluon vertex as a function
  of orthogonal gluon and antiquark momenta.\myhfill]{
		\includegraphics[width=0.40\textwidth]{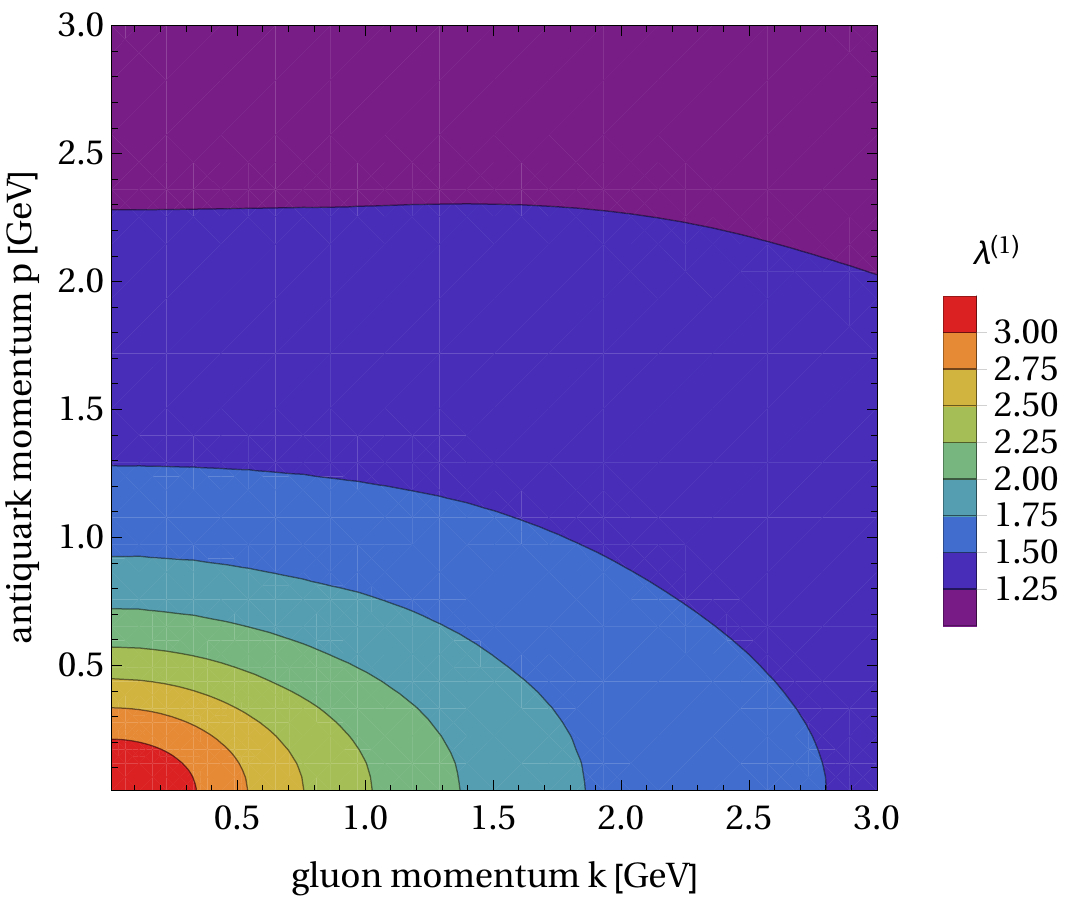}
		\label{fig:quarkgluon_3D}
              } \hfill \subfloat[Dressing functions of four-Fermi
              channels that are not dynamically
              hadronised. Here the same conventions as in 
\cite{Mitter:2014wpa}
	      have been used for labelling the dressing 
functions.\myhfill.]{\includegraphics{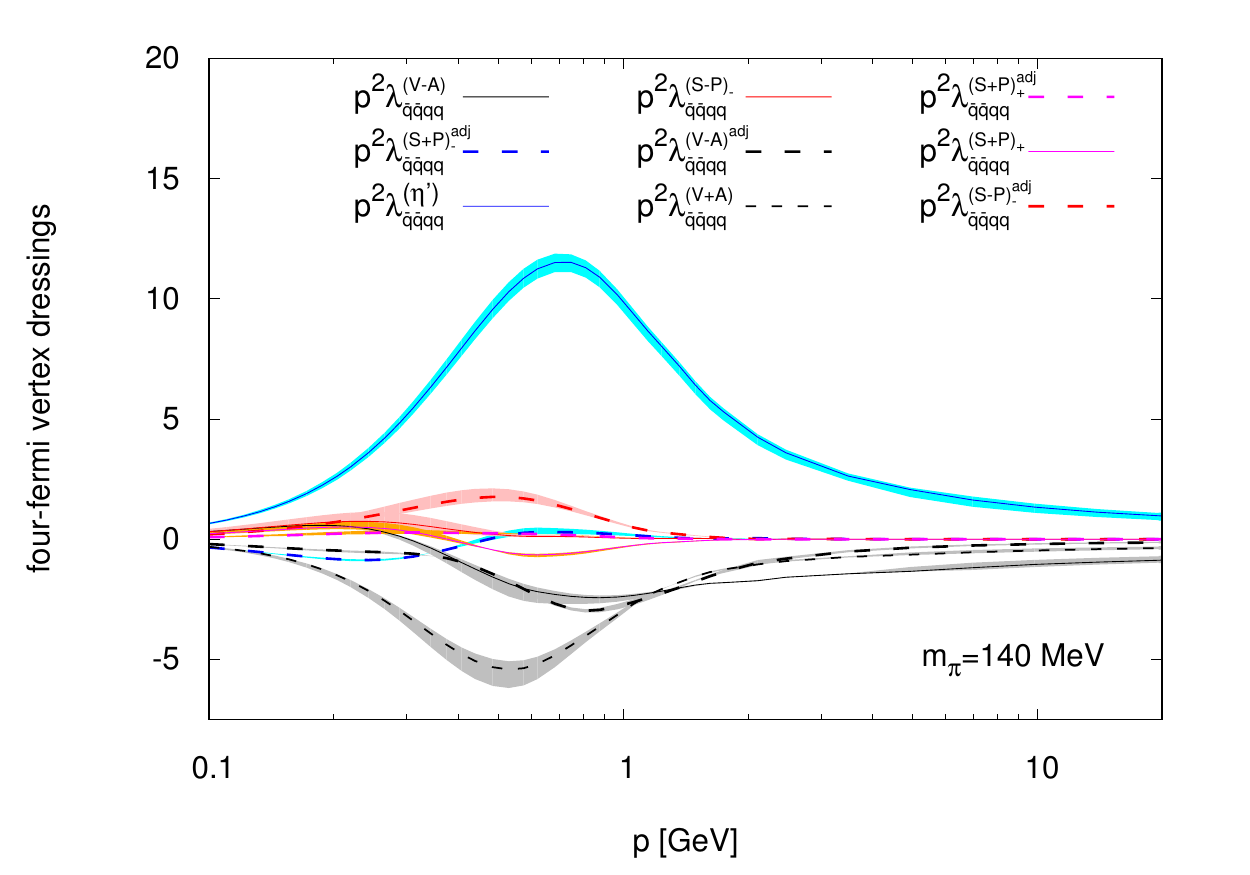}
		\label{fig:fourfermi}}
	\caption{Quark-gluon and four-Fermi vertices.}
	\label{fig:quarkglue}
\end{figure*}

The channels of the four-Fermi interaction that are not dynamically
hadronised are shown in \Fig{fig:fourfermi} at the u-channel momentum
configuration. Clearly, all of these channels remain finite on these
Euclidean momentum configurations. Since the poles that would
correspond to the respective bound-state masses are too far from the
investigated Euclidean momentum configurations, no conclusions about
the spectrum can be drawn at this stage, see \cite{Rennecke:2015eba}
for an investigation where the vector channels have been dynamically
hadronised as well.

At the symmetric point, a quark-gluon running coupling can
be extracted from the vertex dressing via \eq{eq:runcoupqgv}. It
is shown along with the gluonic running couplings in \Fig{fig:running_coupling}.
The Slavnov-Taylor identity for the quark-gluon vertex
with a trivial quark-ghost scattering kernel implies a deviation of the
quark-gluon running coupling from the pure glue running couplings
in the \mbox{(semi-)perturbative} regime.
Only by including
quantum corrections to the quark-ghost scattering, \cf
 \cite{Davydychev:2000rt,Aguilar:2010cn,Rojas:2013tza,Aguilar:2014lha,Aguilar:2016lbe},
\Sec{sec:STI} and \App{app:quarkghostscatteringkernel}, the corresponding quantum
corrections to the ghost-gluon vertex are compensated and degeneracy
of all the different running couplings is restored.
The quark-gluon running coupling shown in \Fig{fig:running_coupling}
is the most important ingredient for a quantitatively and even qualitatively correct
description of spontaneous chiral symmetry breaking. In particular, the range
of momenta, where it exceeds the critical value of the coupling 
$\alpha_{\rm cr}\approx 0.86$ determines, if, and 
to which extent, chiral symmetry breaking occurs \cite{Braun:2011pp,Mitter:2014wpa}. 
Consequently, a precise
determination of the quark-gluon running coupling is of utmost importance, since the
corresponding error directly translates into the quark mass function, as can
bee seen by comparing the bands in 
\Figs{fig:running_coupling}{fig:quark_prop_mpi}.
Although the resummation scheme, and as a consequence also the 
mechanism for chiral symmetry breaking, is different, an analogous 
sensitivity on the quark-gluon interaction strength is also found in 
Dyson-Schwinger studies \cite{Hopfer:2013np}. 

Further results on higher-order vertex functions are presented in
\Fig{fig:quark_gluon_int} and are only discussed briefly at this
point.  Turning to higher quark-gluon interaction vertices, we want to
highlight the fact that this work incorporates the first direct
computation of these interactions. Already earlier investigations
found clear evidence for their quantitative importance
\cite{Mitter:2014wpa}, but inferred their value only indirectly from
the quark-gluon-vertex, exploiting the idea of an expansion in terms
of BRST-invariant operators $\bar q\slashed{D}^n q$. Here we still use
this idea as an organising principle for the basis construction, see
\App{app:quarkvertices}.  The crucial improvement in comparison to
\cite{Mitter:2014wpa} is the direct calculation of the corresponding
dressing functions and their self-consistent back-coupling into the
system of equations at the symmetric momentum configuration.
In comparison to the approximation used in \cite{Mitter:2014wpa},
the directly calculated two-quark-two-gluon vertex leads to a 
moderately enhanced quark-gluon vertex.
Exemplary results for the dressing functions of the
two-quark-two-gluon vertex are shown in
\Fig{fig:twoquarktwogluon}-\Fig{fig:twoquarktwogluon_3}.  Furthermore
we present also exemplary results for the leading tensors in the
two-quark-three-gluon vertex, see
\Fig{fig:twoquarkthreegluon}. However, they turn out to be of
sub-subleading importance for the overall system of correlation
functions. These results are complemented by results on quark-meson
interaction vertices in the soft-pion channel, see
\Fig{fig:yukawa}. For the classical tensor
$\mathcal{T}_{\bar q q \pi}^{(1)}$, see \eq{eq:barqpiq_tensors}, this
is the momentum-channel that is relevant to the momentum-dependent
dynamical hadronisation procedure discussed in \App{app:dyn_had}.
Although of sub-subleading importance for the system of equations, the
other momentum-dependent tensor structures of the quark-meson Yukawa
interaction shown in \Fig{fig:yukawa} are important ingredients in 
bound-state
studies, see \eg
\cite{Roberts:1994dr,Alkofer:2000wg,Eichmann:2016yit}. The same
applies to higher-order quark-pion scattering operators, also resolved
momentum-dependently and depicted in \Fig{fig:twoquarknmeson}.

\begin{figure*}[t]
	\centering 
	\subfloat[Running couplings as defined in \eq{eq:runcoup} and \eq{eq:runcoupqgv}.
	The inset shows the relative
	deviations
	$\Delta X=(\alpha_{\bar c c A}-\alpha_{X})/\alpha_{\bar cc A}$
	compared to the ghost-gluon vertex running coupling in the
	semi-perturbative regime. The abscissa of the inset is identical
	to the abscissa of the full plot.
	\myhfill]{
		\includegraphics{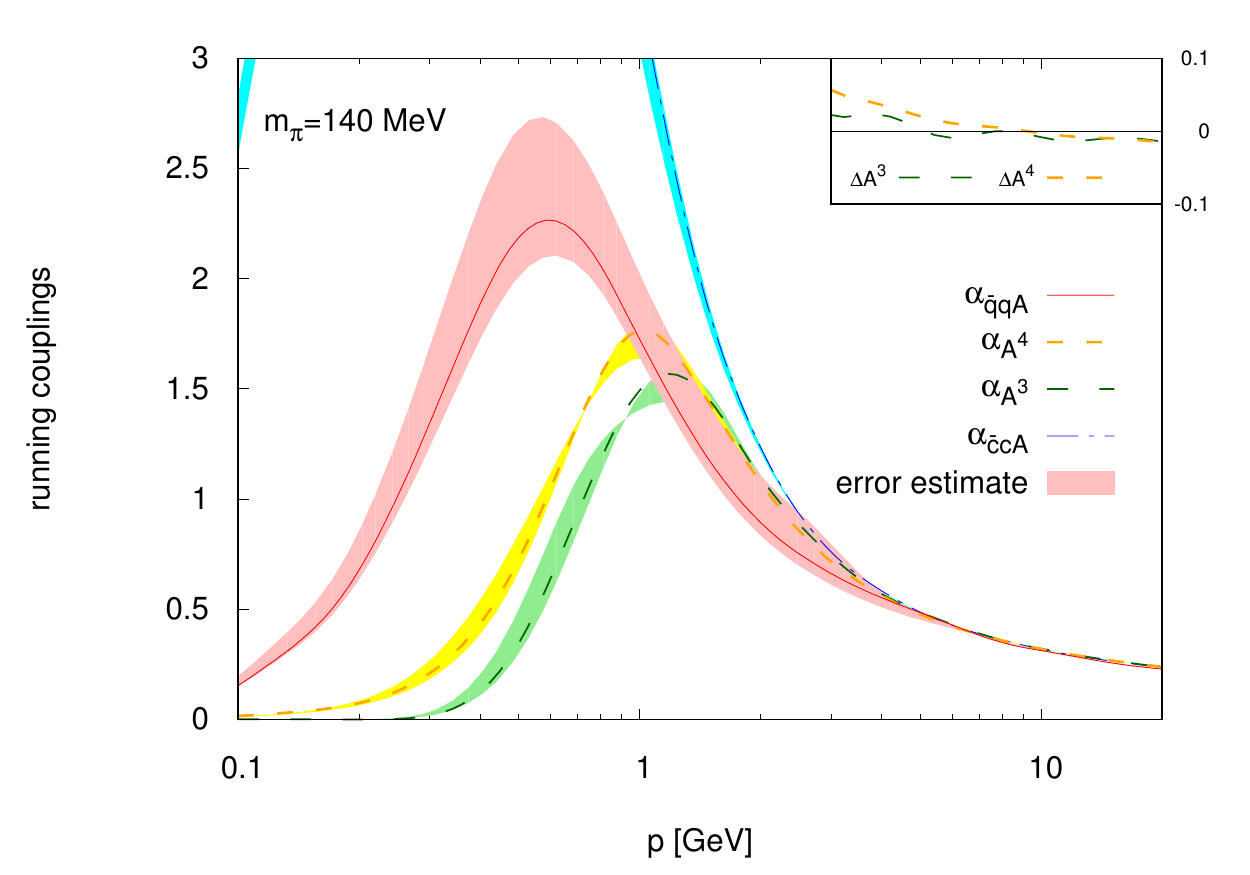}
		\label{fig:running_coupling}} \hfill \subfloat[Quark
              propagator dressing functions, $M_q(p)$ (lower line, in
              GeV) and $1/Z_q(p)$ (upper line) for different
              truncations in comparison to lattice data
              \cite{Oliveira:2016muq, *Sternbeck:PC17}. Here, our results have a pion mass of $m_\pi \approx \SI{285}{\MeV}\,$.
              \myhfill]{
              	\includegraphics{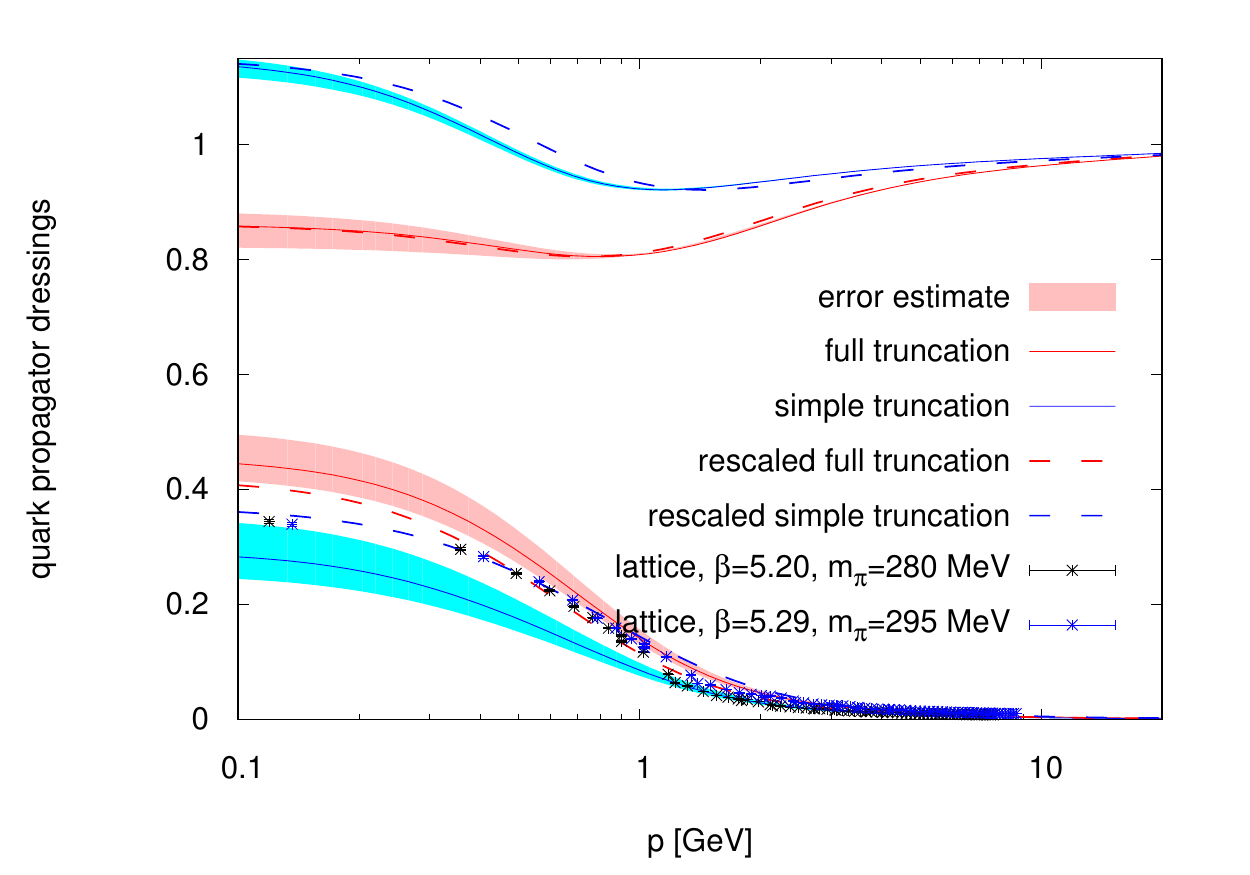}
           \label{fig:truncation_test}}
	\caption{Running couplings and truncation dependence of the quark propagator.}
	\label{fig:main_result2}
\end{figure*}

\section{Discussion}
\label{sec:AppCon} 

One of the main conclusions we can draw from the results presented in
the previous section is that a quantitatively reliable description of
spontaneous chiral symmetry breaking within the functional methods
requires very precise results on the quark-gluon interaction
$\Gamma^{(3)}_{\bar q q A}(p,q)$, see also \cite{Hopfer:2013np} for
corresponding observations in the DSE framework.
Additionally, we find that the transverse running couplings defined 
from the different vertices deviate considerably from each other at momenta
around and below the scale of QCD. We interpret this as a clear signal for 
the non-applicability of the respective Slavnov-Taylor identities to constrain 
the transversely projected vertices in this regime.
Furthermore, we observe that it seems to be very hard to find truncations that lead to
a consistent (semi-)perturbative running of the coupling strengths of the matter and gauge
sectors.  To the best of our knowledge, there are two sources for this
scale separation.  On the one hand, a precise classification of
vertices and diagrams in loop orders is difficult to achieve within
the functional approaches.  As a result, the two sectors might run
with different loop orders in the semi-perturbative regime.
Furthermore, the chosen truncation, and in particular the momentum
dependencies of the vertex dressing functions, might violate the BRST
symmetry of the different sectors to a different degree.

These findings emphasise the need for truncations that lead to a 
consistent running of all of its constituents. 
We find this to be a considerably harder task in fully unquenched QCD,
as compared to the gauge sector, where consistent running was 
already achieved in \cite{Cyrol:2016tym}.
In particular, the STIs allow for consistency checks of the running
of the different correlators. However, this requires the computation
of their longitudinal as well as transverse parts. 
To this end, also other functional relations such as \eg 
Dyson-Schwinger (DSE), 2PI
relations or transverse Ward-Takahashi identities, 
(see \App{app:tWTI} and 
\cite{Takahashi:1985yz,Kondo:1996xn,He:2000we,Qin:2013mta,Aguilar:2014lha}) can be 
employed. Such an elaborate approach will be discussed elsewhere

In this work, we tackled these issues with a two-step strategy. First,
we used the Slavnov-Taylor identity for the quark-gluon vertex, 
\cf \Sec{sec:STI}, to constrain the
perturbative behaviour of its transversely projected classical tensor structure.  In
particular, we find that using this STI forces the degeneracy of the
running couplings of the matter and glue sector in the perturbative
regime.  Second, we extended the truncation to include higher
quark-gluon interactions, namely the two-quark-two-gluon and the
two-quark-three-gluon $1$PI correlators, $\Gamma^{(4)}_{\bar q q A^2}$
and $\Gamma^{(5)}_{\bar q q A^3}\,$, with a consistent set of basis
elements.  This allowed us to calculate the non-perturbative features
of the quark-gluon interaction with unprecedented precision.

Consequently, we have two tools at our disposal to assess the
systematic error.  First, we vary the transition scale up to which
the STI \eq{eq:STI} is used to constrain the quark-gluon vertex, \cf
the discussion in \Sec{sec:STI}. In line with the reasoning to
determine $\LambdaSTI\,$, see \Sec{sec:STI}, we observe sizeable
deviations from the one-loop STI in the classical vertex structures of
the gluonic vertices, below \SIrange{3}{5}{\GeV}, \cf the inlay in
\Fig{fig:running_coupling}.  Therefore, we vary the transition scale
from STI-constrained to fully-calculated vertex, $\LambdaSTI\,$, between
$3$ and \SI{7}{\GeV} to obtain the shown bands. Our main results
(solid lines) are obtained with a transition scale of \SI{5}{\GeV}.

Second, we compare our best result to results obtained within simpler
truncations of the matter sector. In \Fig{fig:truncation_test} we show
a comparison of the corresponding quark propagators. Here, the blue
result has been obtained in a truncation, where only the classical
tensor structure for the quark-gluon vertex has been taken into
account on a symmetric momentum configuration, similarly to the
approximation of the vertices in the glue sector.  The difference
between the blue and red result gives an estimate for the upper bound
of the truncation error.

Finally, we want to point out that the difference between the bands in
\Fig{fig:truncation_test} makes the resulting error look worse than it
will actually be in applications to the phase structure and bound
state spectrum of QCD.  In such investigations, we would have to set
the scale of the theory in terms of observables like the pion decay
constant or the quark condensate. Simulating this procedure by using
the value of the constituent quark mass $M_q(p)$ at $p=\SI{0.5}{\GeV}$ to 
set this scale, we obtain the dashed curves in \Fig{fig:truncation_test}
for the two truncations. The difference between the resulting quark
mass functions gives a more realistic estimate of the truncation
effects on observables, since only relative effects will be important
in this case.

\section{Summary and Conclusions}
\label{sec:summary}

In this work we investigated correlation functions in unquenched Landau gauge
QCD with $N_f=2$ quark flavours. 
This analysis was performed within the Functional Renormalisation Group
approach in a vertex expansion scheme for the effective action. We presented
a self-consistent solution for the hitherto largest system of correlation 
functions aiming at quantitative precision. The numerical results for 
gluon propagator and quark mass function were found to be in very good agreement 
with corresponding lattice results.
Results for the propagators with different values of the pion
mass were presented.  In particular, this includes results with small pion 
masses, which are notoriously difficult to
obtain in lattice simulations. Finally, we also showed results for the
higher quark-quark, quark-gluon and quark-meson interactions that are
part of our truncation.

Special emphasis was put on the importance of the correct running of
vertices.  In particular, the STI-consistent running of the
quark-gluon vertex was found to be of utmost importance for
the qualitative and quantitative description of
chiral symmetry breaking.
Although our truncation is of unprecedented extent, we still observe
a mismatch in the scales of the glue and the matter sector.
Therefore, the quark-gluon vertex STI was used to guarantee the correct running in the
(semi-)perturbative regime, whereas the full non-perturbative structure
of the vertex was numerically calculated at non-perturbative momenta.
In our truncation, we found that the STIs imply degeneracy of the running couplings
defined from different vertices only if quantum corrections to the quark-ghost 
scattering kernel are included.

These results provide a major milestone towards the goal of first-principle
investigations of the phase structure of QCD. Having established a stable
truncation that allows to solve the FRG equations without modelling input, 
investigations
at finite temperature are now the next logical step. In parallel, further investigations
of the stability of the truncation will be of crucial importance.
These further tests and improvements of the 
truncation will be particularly important for investigations at finite densities,
which require the full quantitative control over the fluctuation degrees of freedom.

\acknowledgments
The authors thank C.~Aguilar, R.~Alkofer,
D.~Binosi, J.~Braun, G.~Eichmann, W.-J.~Fu, C.~Fischer,
M.~Huber, M.~Leonhardt, J.~Papavassiliou, C.~Roberts, A.~Sternbeck, 
and R.~Williams for discussions.

This work is supported by EMMI,
the grant ERC-AdG-290623, the FWF through
Erwin-Schr\"odinger-Stipendium No.\ J3507-N27, the Studienstiftung des
deutschen Volkes, the DFG through grant STR 1462/1-1, the BMBF grant
05P12VHCTG, and is part of and supported by the DFG Collaborative
Research Centre "SFB 1225 (ISOQUANT)". It is also supported in part by the Office of Nuclear
Physics in the US Department of Energy's Office of Science 
under Contract No. DE-AC02-05CH11231.

\appendix

\begin{figure*}
	\subfloat[Two-quark-two-gluon vertex $\Gamma^{(4)}_{\bar q q A^2}(\psym)$, symmetric tensors.]
	{
		\includegraphics{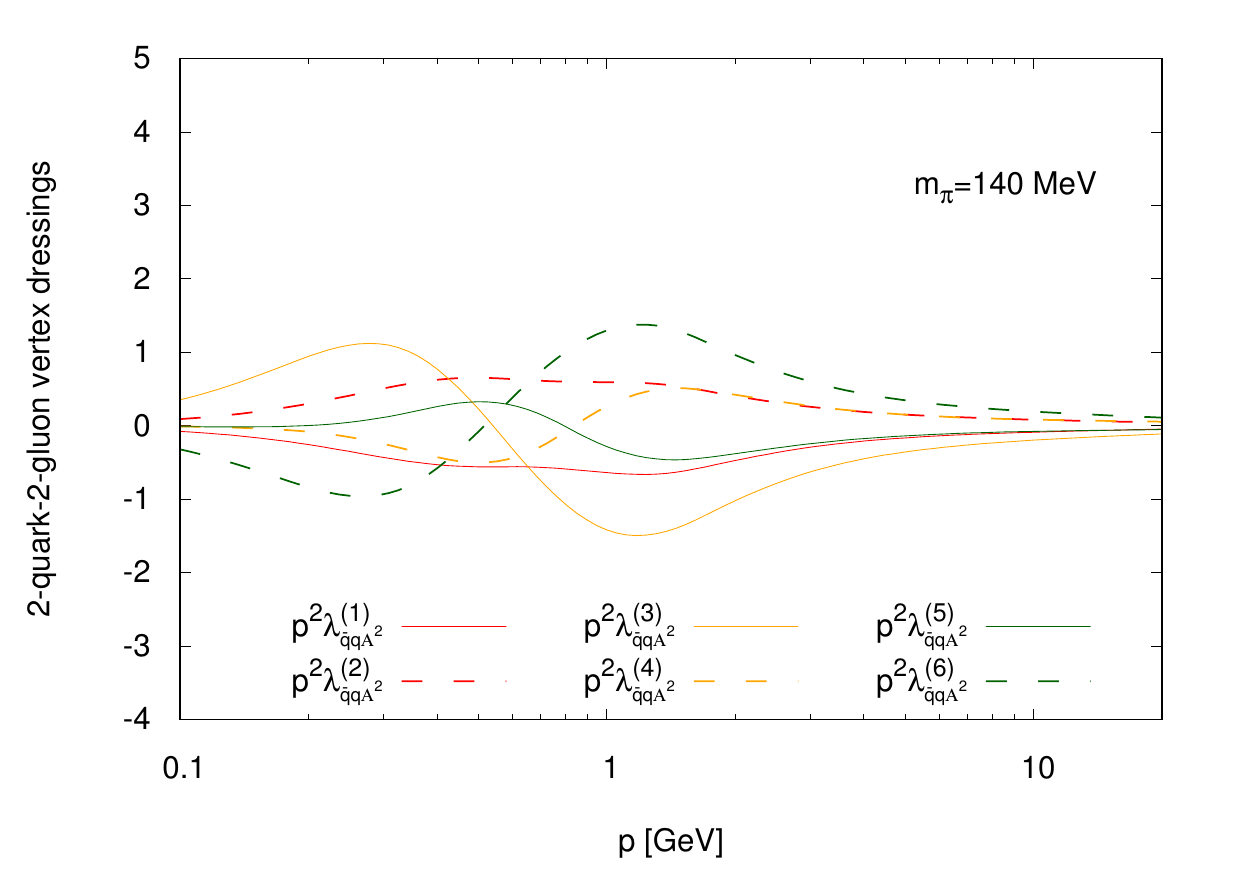}
		\label{fig:twoquarktwogluon}
	}
	\hfill
	\subfloat[Two-quark-two-gluon vertex $\Gamma^{(4)}_{\bar q q A^2}(\psym)$, symmetric tensors.]
	{
		\includegraphics{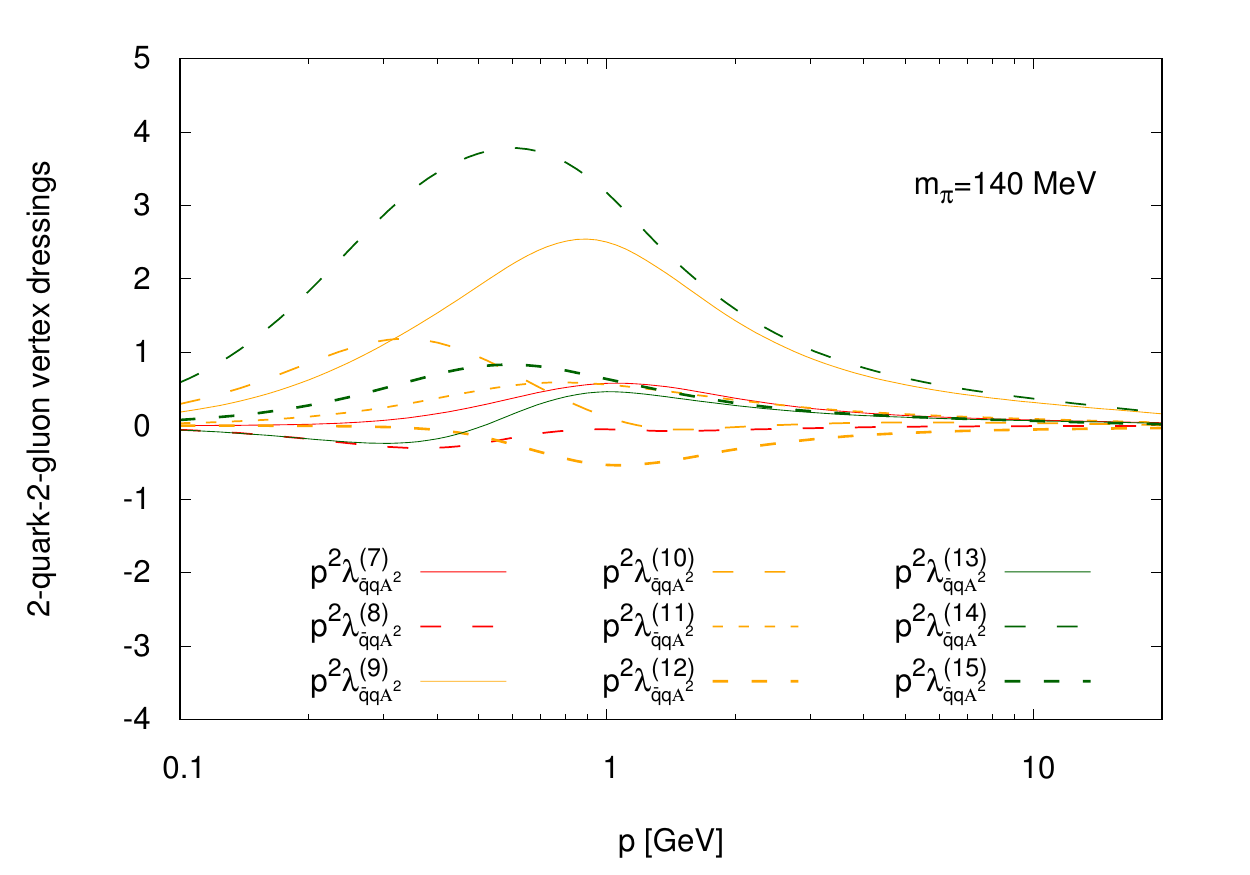}
		\label{fig:twoquarktwogluon_2}
	}
	\\
	\subfloat[Two-quark-two-gluon vertex $\Gamma^{(4)}_{\bar q q A^2}(\psym)$, symmetry-breaking tensors.]
	{
		\includegraphics{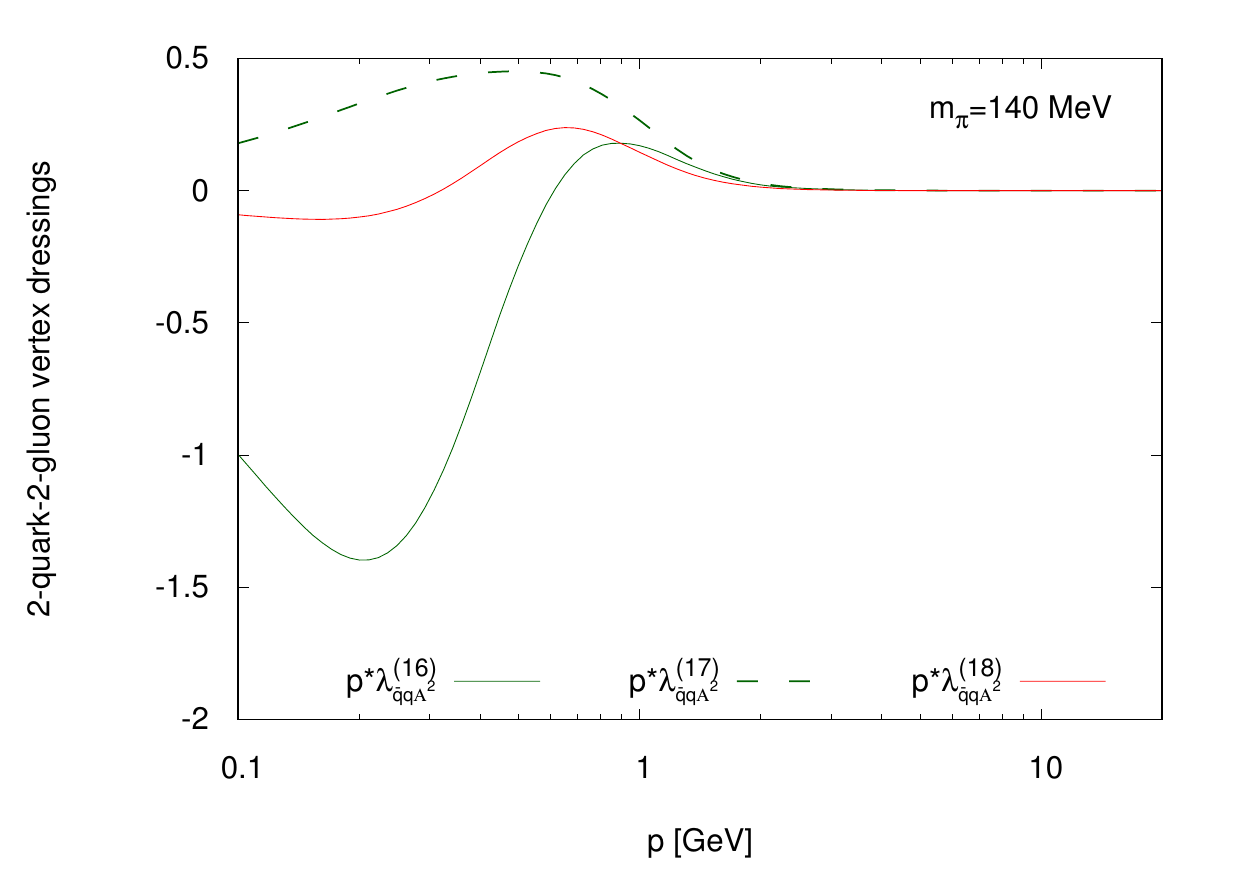}
		\label{fig:twoquarktwogluon_3}
	}
	\hfill
	\subfloat[Two-quark-three-gluon vertex $\Gamma^{(5)}_{\bar q q A^3}(\psym)\,$.]
	{
		\includegraphics{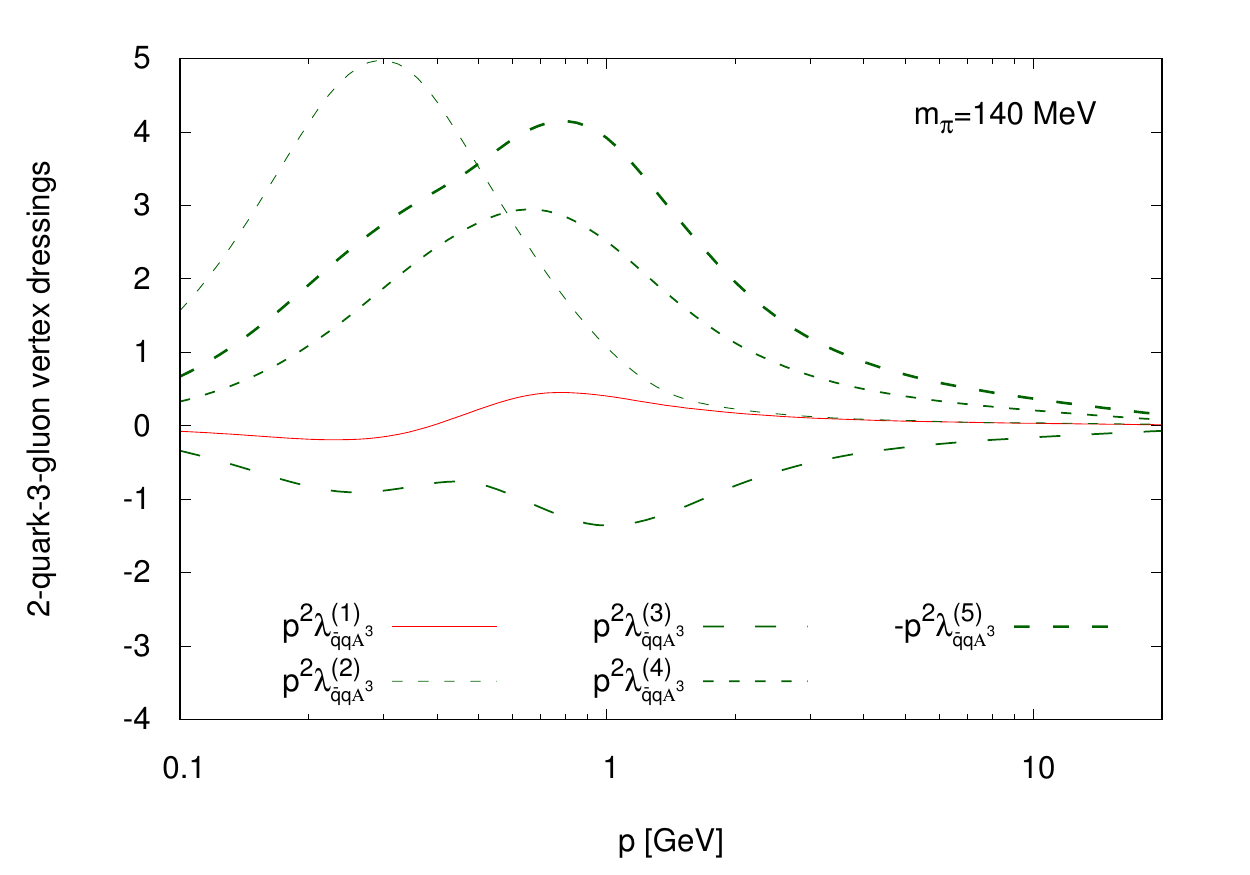}
		\label{fig:twoquarkthreegluon}
	}
	\\
	\subfloat[Yukawa interactions $\Gamma^{(3)}_{\bar q q \phi}(p,-p)\,$.]
	{
		\includegraphics{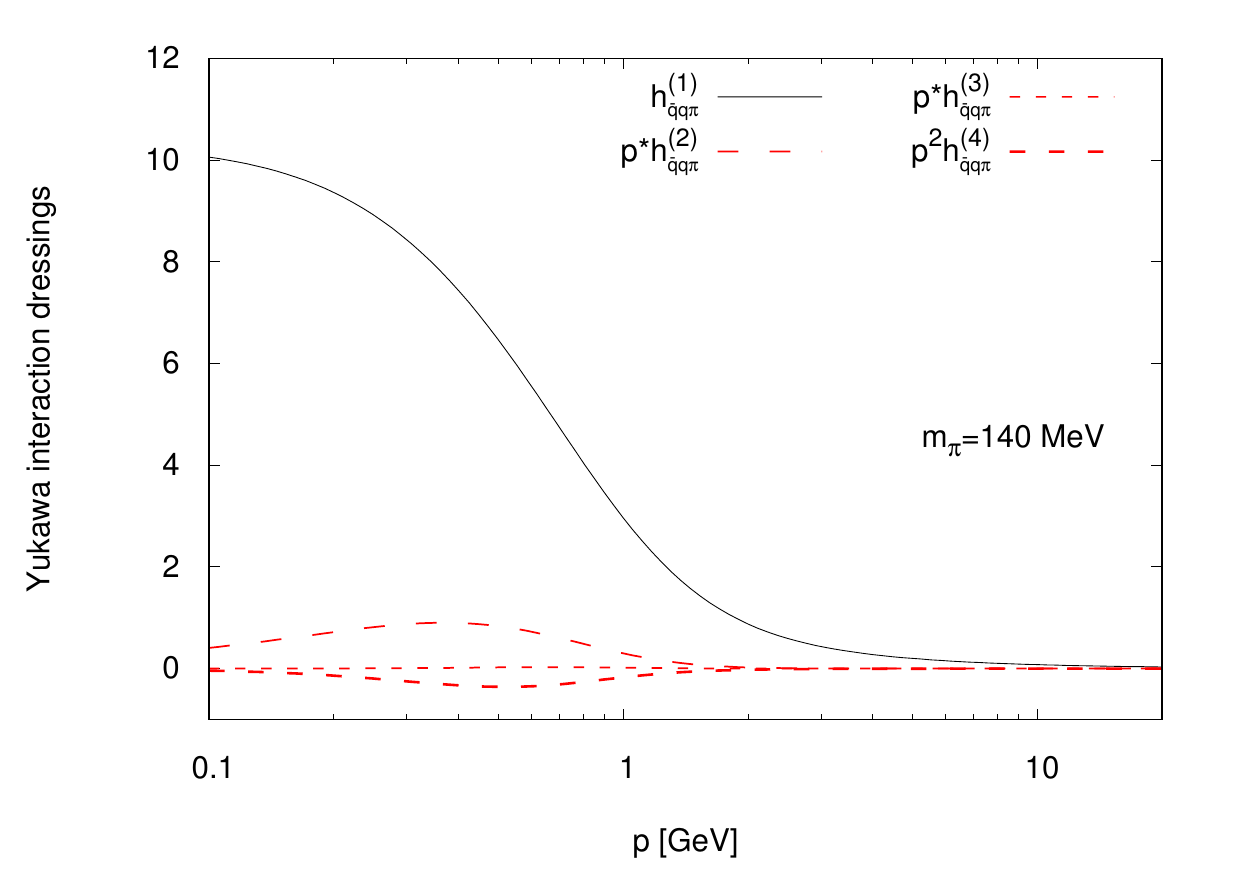}
		\label{fig:yukawa}
	}
	\hfill
	\subfloat[Two-quark-$n$-meson vertex $\Gamma^{(2+n)}_{\bar q q \phi^n}(p,-p,\vec{0})\,$.]
	{
		\includegraphics{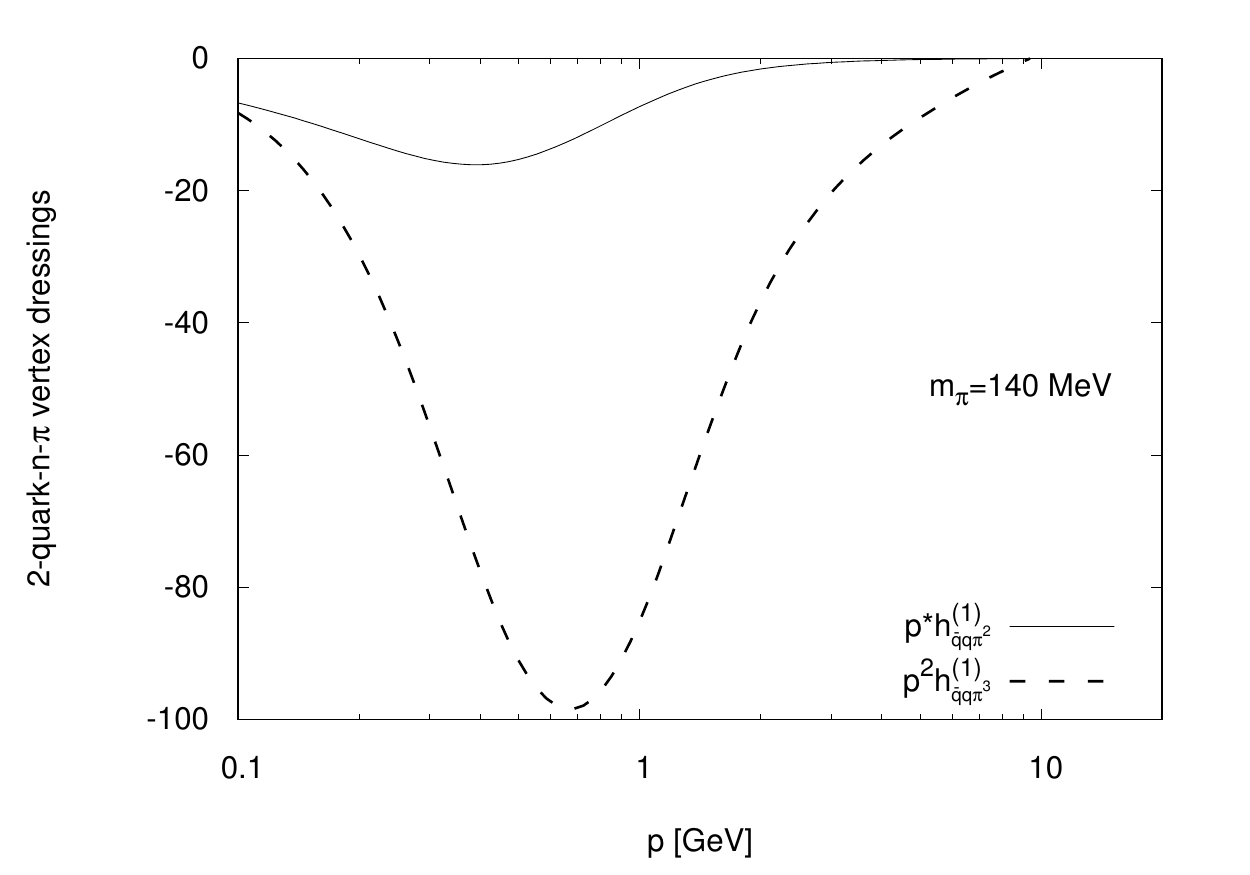}
		\label{fig:twoquarknmeson}
	}
	\caption{
		Dressing functions of the two-quark-n-gluon and two-quark-n-meson interactions defined in \App{app:quarkvertices}.
		All shown dressing have been made dimensionless by multiplication with according powers of the average momentum.
		\myhfill
	}
	\label{fig:quark_gluon_int}
\end{figure*}

\section{Momentum-dependent dynamical hadronisation}
\label{app:dyn_had}

In the resummation scheme defined by the FRG, chiral symmetry breaking
is driven by the four-Fermi interaction, which in turn is created
dynamically from box diagrams with a two-gluon exchange, see 
\eg \cite{Braun:2011pp}. In the case
of a momentum-independent approximation, the spontaneous breaking of
chiral symmetry is signalled by a divergence in the pion channel of
the four-Fermi interaction. As a consequence, divergences appear also
in the other channels. The emergence of this singularity is a
consequence of the emerging pion pole, whose proper description
requires momentum dependencies.  In order to include the missing
momentum dependencies as efficiently as possible and to be able to access
the symmetry-broken regime, the dynamical hadronisation technique is
applied
\cite{Gies:2001nw,Pawlowski:2005xe,Floerchinger:2009uf,Mitter:2014wpa,Braun:2014ata}.
Once the auxiliary mesonic field variables are introduced, the remaining
channels of the four-Fermi vertex remain finite at all finite
RG scales, see \Fig{fig:fourfermi}.

Analogous to \cite{Pawlowski:2005xe,Mitter:2014wpa,Braun:2014ata} we
introduce a scale-dependent dynamically hadronised field in the
scalar-pseudoscalar channel of the four-Fermi interaction by defining
the scale derivative of its field expectation value
\begin{equation}
	\label{eq:rebosfield}
	\dot\phi_k^a(p)=\int_q \partial_t\rebosA(p-q,q) \left[\bar q \flavT{a} q\right](p-q,q)\,.
\end{equation}
Here, the dot indicates the derivative with respect to
$t=\log(k/\Lambda)\,$, the $\flavT{a}$, $a\in \{1,2,3\}\,$, correspond to the
Pauli matrices divided by $2$ and $\flavT{0}$ is the unit matrix divided by
$2$.  Therefore, $\phi_k^a$ represents a bosonic field with the
quantum numbers of the pions ($f_0(500)$) for $a=1,2,3$ ($a=0$). The
main difference to the procedure used previously in
\cite{Pawlowski:2005xe,Mitter:2014wpa,Braun:2014ata} is the momentum
dependence, \ie $\partial_t \rebosA(p-q,q)\,$, and the absence of an
additive term $\partial_t \rebosB(p)\,\phi_k(p)$ on the right hand side of
\eq{eq:rebosfield}. Such a term simply introduces a momentum-dependent
re-scaling of the wave function renormalisation $Z_{\phi,k}(p)$ of 
$\phi\,$,
and is hence not considered here. 

The introduction of $\phi_k^a$ leads to an additional term in the
standard flow equation, which becomes
\begin{align}
	\label{eq:DynHad}
	\dot\Gamma_k[\Phi] = \frac{1}{2}\,\Tr\, 
	\frac{1}{\Gamma^{(2)}_k+R_k}\dot R_k - \frac{\delta \Gamma_k}{\delta \phi^a} \dot \phi_k^a\,.
\end{align}
Consequently, any $n$-point function that includes at least one 
quark-antiquark pair $\bar q\flavT{a} q$ gets an additional contribution,
\begin{align}
	\label{eq:DynHadexplicit}
	&\Delta\dot\Gamma^{(n)}_{\bar q \flavT{a} q\varphi_3\cdots\varphi_n}(p_1,p_2,\dots,p_{n}) 
		= \eqnewline
	&\qquad-\frac{\delta^{n}\left( \frac{\delta \Gamma_k}{\delta
		\phi^a}\partial_t \rebosA\left[\bar q \flavT{a}
		q\right]\right)}{\delta\varphi_n(p_n)\cdots\delta\varphi_3(p_3)\delta[\bar q\flavT{a}
		q](p_1,p_2)}\,,
\end{align}
where the integration over momenta in the numerator is
implicit. Therefore, the flow of any $n$-point function
$\Gamma^{(n)}_{\bar q q\varphi_3\cdots\varphi_n}\,$, whose combined
quantum numbers of $\varphi_3\dots\varphi_n$ correspond to one of the pions
or the sigma-meson, is modified by the introduction of the
scale-dependent dynamical hadronisation fields.

In particular, the flow of the four-Fermi interaction channel
corresponding to pion or sigma-meson exchange and the according
quark-meson Yukawa interaction are modified as
\begin{align}\label{eq:DynHadveryexplicit}
 &\Delta\dot\Gamma^{(4)}_{(\bar q\flavT{a} q)^2}(p_1,p_2,p_3)
=  \nonumber\\&\qquad\qquad\qquad-\frac{\delta^{n}\left( \frac{\delta \Gamma_k}{\delta \phi^a}\partial_t \rebosA\left[\bar q \flavT{a} q\right]\right)}{\delta[\bar q\flavT{a} q](p_3,p_4)\delta[\bar q\flavT{a} q](p_1,p_2)}\,,
\eqnewline
 &\Delta\dot\Gamma^{(3)}_{(\bar q\flavT{a} q)\phi^a}(p,q) =  - \partial_t \rebosA(p,q)\Gamma_{\phi^a\phi^a}(p+q)\,,
\end{align}
where $p_4=-p_1-p_2-p_3\,$.
Since $\partial_t \rebosA$ is a function of two momenta, we can choose it such that particular momentum
channels of the four-Fermi interaction are rewritten in terms of the exchange of mesons 
represented by the field $\phi^a$ by demanding 
\begin{align}\label{eq:DynHadcondition}
 \text{Flow}\left[\Gamma^{(4)}_{(\bar q \flavT{a} q)^2}\right]  + \Delta\dot\Gamma^{(4)}_{(\bar q \flavT{a} q)^2}\bigg\vert_{\Omega} \equiv 0
\end{align}
on a subset of momenta $\Omega\,$. Dynamically hadronising the momentum 
channel
$\Gamma^{(4)}_{(\bar q \flavT{a} q)^2}(p,q,-p,-q)$ corresponds then to the choice
\begin{align}\label{eq:Adot}
 \partial_t \rebosA (p,q) &= -2\frac{\dot \lambda^{(\pi)}_{\bar q\bar q qq}(p,-q,-p,q)}{h^{(1)}_{\bar q q \phi}(q,p)}\\
&+\frac{\dot \lambda^{(\pi)}_{\bar q\bar q qq}(q,q,-q,-q) h^{(1)}_{\bar q q \phi}(p,-p)}{2h^{(1)}_{\bar q q \phi}(q,p)h^{(1)}_{\bar q q \phi}(q,-q)}\nonumber\\&+\frac{\dot \lambda^{(\pi)}_{\bar q\bar q qq}(p,p,-p,-p) h^{(1)}_{\bar q q \phi}(q,-q)}{2h^{(1)}_{\bar q q \phi}(q,p)h^{(1)}_{\bar q q \phi}(p,-p)}\ ,\nonumber
\end{align}
where $\lambda^{(\pi)}_{\bar q\bar q q q}=\lambda^{(S-P)_+}_{\bar q\bar q q q}+\lambda^{(S+P)_+}_{\bar q\bar q q q}$
with the conventions of \cite{Mitter:2014wpa} and $h^{(1)}_{\bar q q \phi}$ are the dressing functions of the four-Fermi interaction and quark-meson Yukawa interaction respectively.

These equations are simplified considerably in the u-channel rebosonisation, where both (anti-)quarks carry the momentum $(-)p$, leading to
\begin{align}
\label{eq:DynHaduchannel}
 \Delta\dot h^{(1)}_{\bar q q \phi}(p,-p) &= \frac{\Gamma_{\phi^a\phi^a}(0)}{h^{(1)}_{\bar q q \phi}(p,-p)}\dot \lambda^{(\pi)}_{\bar q\bar q qq}(p,p,-p,-p)\ .
\end{align}
This channel is particularly interesting, because the quark mass function receives an analogous correction 
from this channel by multiplying the above equation with the expectation value of $\langle\phi^0\rangle$. 
Therefore, this is the momentum channel that is bosonised in this work.
Finally, \eq{eq:DynHaduchannel} reduces to the well-know result 
\cite{Gies:2001nw,Braun:2014ata},
\begin{align}
 \Delta\dot h^{(1)}_{\bar q q \phi} &= 
	\frac{m_\pi^2}{h^{(1)}_{\bar q q \phi}}\dot
		\lambda^{(\pi)}_{\bar q\bar q qq}\,,
\end{align}
at vanishing momenta.

\section{Truncation scheme}
\label{app:truncationscheme}

As discussed in \Sec{sec:vertexexpansion}, we classify all tensors into 
classical, leading non-classical and sub-leading non-classical and neglect all
sub-sub-leading contributions. Although previous results indicate that the 
importance of different constituents of the truncation might be connected to 
BRST-invariant operators \cite{Mitter:2014wpa}, we perform additional,  
explicit checks to test the importance of different parts of our
truncation. The identification of the classical tensors is clear,
where we additionally interpret the Yukawa interaction between quarks
and mesons as well as the meson propagators as classical tensors. The
latter are present in our truncation because of a momentum-dependent
version of the dynamical hadronisation technique
has been used to represent the leading channel of the dynamically
created four-quark interaction in terms of exchange of mesons,
see \cite{Gies:2001nw,Pawlowski:2005xe,Floerchinger:2009uf,Mitter:2014wpa,Braun:2014ata} 
and \App{app:dyn_had}. 
The resulting classification of the different vertices and their
tensor structures considered in our truncation is summarised in \Tab{tab:assignment}.

The remaining equations are still very large and we deviate in a controlled manner 
from the above expansion scheme in some equations in order to make the system of 
equations numerically tractable.
\begin{itemize}
 \item First, we completely neglect the sub-leading non-classical quark-gluon vertex tensors, 
which has been checked explicitly to be a very good approximation, see also
\cite{Hopfer:2013np,Williams:2014iea,Mitter:2014wpa,Williams:2015cvx}.
 \item Second, we ignore any sub-leading non-classical contributions to the four-gluon
vertex. Since this vertex is the least important of the classical vertices, we expect
this to be a good approximation, although explicit checks are amiss due to the size
of these equations.
 \item Third, contributions from the tensor ${\cal T}^{(4)}_{\bar q qA}$ are 
ignored in the equation for the dressing of ${\cal T}^{(7)}_{\bar q qA}$ 
and vice versa, which has been tested to be a very good approximation. 
 \item Fourth, we include the effect of the two-quark-three-gluon vertex in the 
leading non-classical tensors of the quark-gluon as well as the 
two-quark-two-gluon vertex, which has been found to yield only small 
corrections.
 \item Fifth, we ignore the effect of ${\cal T}^{(4)}_{\bar q qA}$ 
and ${\cal T}^{(i)}_{\bar q qA^2}\,$, 
$i\in\{1,\dots,6,8,9,10,12,13,16,17,18\}\,$, in the 
equation for the dressings of ${\cal T}^{(i)}_{\bar q qA^2}\,$, 
$i=\{1,\dots,15\}$ and furthermore the effect of
${\cal T}^{(7)}_{\bar q qA}$ and
${\cal T}^{(i)}_{\bar q qA^2}\,$, $i\in\{1,\dots,15\}\,$, in the 
equation for the dressings of ${\cal T}^{(i)}_{\bar q qA^2}\,$, 
$i\in\{16,17,18\}\,$. This approximation has been explicitly
checked to be very good. 
 \item Sixth, we always feed back all purely mesonic 
interactions, which is particularly important for the effective potential 
and the mesonic propagators.
\end{itemize}
 
\begin{table}
	\centering
	{\renewcommand{\arraystretch}{1.5}
	\begin{tabular}{c|c|c|c}
		object & classical & leading & sub-leading\\\hline
		$\Gamma^{(2)}_{A^2}$ & all &  & \\
		$\Gamma^{(2)}_{\bar c c}$ & all &  & \\
		$\Gamma^{(3)}_{\bar c c A}$ &$\mathcal{T}^{(1)}_{\bar c c A}$ & &\\
		$\Gamma^{(3)}_{A^3}$ & $\mathcal{T}^{(1)}_{A^3}$ & &\\
		$\Gamma^{(4)}_{A^4}$ & $\mathcal{T}^{(1)}_{A^4}$ & &\\\hline
		$\Gamma^{(2)}_{\bar q q}$ & all &  & \\
		$\Gamma^{(3)}_{\bar q q A}$ &$\mathcal{T}^{(1)}_{\bar q q A}$ & $\mathcal{T}^{(4)}_{\bar q q A}$, $\mathcal{T}^{(7)}_{\bar q q A}$ & remaining $\mathcal{T}^{(i)}_{\bar q q A}$\\
		$\Gamma^{(4)}_{\bar q q A^2}$ & & $\mathcal{T}^{(1)}_{\bar q q A^2}\ldots\mathcal{T}^{(18)}_{\bar q q A^2}$ & \\
		$\Gamma^{(5)}_{\bar q q A^3}$ & & &$\mathcal{T}_{\bar q q A^3}^{(1)}\ldots\mathcal{T}_{\bar q q A^3}^{(5)}$  \\
		$\Gamma^{(4)}_{\bar q \bar q q q}$&& $\mathcal{T}^{(\pi)}_{\bar q \bar q q q}$, $\mathcal{T}^{(\eta')}_{\bar q \bar q qq}$ & remaining $\mathcal{T}^{(i)}_{\bar q \bar q qq}$ \\
		$\Gamma^{(3)}_{qc Q_q}$&& $\mathcal{T}^{(1)}_{q c Q_q}$, $\mathcal{T}^{(4)}_{q c Q_q}$ & \\\hline
		$\Gamma^{(2)}_{\phi^2}$ &all & &\\
		$\Gamma^{(3)}_{\bar q q \phi}$&$\mathcal{T}^{(1)}_{\bar q q \phi}$&&$\mathcal{T}^{(2)}_{\bar q q \phi}\ldots \mathcal{T}^{(4)}_{\bar q q \phi}$\\
		$\Gamma^{(4)}_{\bar q q \phi^2}$&&&$\mathcal{T}_{\bar q q \phi^2}$\\
		$\Gamma^{(5)}_{\bar q q \phi^2}$&&&$\mathcal{T}_{\bar q q \phi^3}$\\
		$\Gamma^{(n)}_{\phi^n}$& $\mathcal{T}_{\phi^3}$, $\mathcal{T}_{\phi^4}$ && $\mathcal{T}_{\phi^5} \ldots \mathcal{T}_{\phi^{12}}$\\
	\end{tabular}
	}
	\caption{Assignment of considered vertices/tensor structures from \Fig{fig:truncation} to the three classes, \cf \Tab{tab:backcoupling}.}
	\label{tab:assignment}
\end{table}

Finally, we do not calculate the full momentum dependence of all of the
dressing functions that appear in our truncation.
The calculated momentum dependency of each constituent of our truncation
is shown in \Fig{fig:truncation}. Here,
$\psym$ represents the symmetric momentum configuration defined by
$\psym^2 = p_i\cdot p_i = -(n-1)p_i\cdot p_j$ with
$i\neq j\in\{1,\dots,n\}$ for any $n$-point function
$\Gamma^{(n)}_{\Phi_1\dots\Phi_n}(p_1,\dots,p_{n-1})\,$. The momentum
dependence on this symmetric configuration is then used to evaluate
the momentum dependence on any other momentum configuration via the
approximation
$\Gamma^{(n)}_{\Phi_1\dots\Phi_n}(p_1,\dots,p_{n})\approx
\Gamma^{(n)}_{\Phi_1\dots\Phi_n}(\psym(p_1,\dots,p_{n}))$
with $\psym(p_1,\dots,p_{n})=\sqrt{(p_1^2+\dots+p_n^2)/n}\,$. A similar
approximation is used to calculate the full momentum dependencies from
the calculated reduced momentum dependencies of the quark-meson
interactions, \cf \App{app:twoquarknmeson}. Finally, the mesonic
interactions are approximated as momentum-independent and calculated
at vanishing momentum.

In comparison to the approximation used in \cite{Cyrol:2016tym}, we have
ignored additional momentum dependencies in the pure gauge sector of
the theory due to the computational costs of taking these into
account.  The effect of this approximation is an overestimation of the
bump in the gluon propagator of \SIrange{5}{10}{\percent} as shown 
explicitly in
\cite{Cyrol:2016tym}.  On the other hand, first exploratory checks indicate
an underestimation of the bump in the gluon propagator of 
\SI{10}{\percent}
due to our restricted momentum-dependence of the two-quark-two-gluon
vertex in the quark-tadpole of the gluon-propagator equation. 
Consequently, the net effect of these
approximations is expected to be small. This will be checked, together
with the influence of the neglected additional momentum dependencies, 
tensors and non-classical correlation functions in future investigations.

\section{Quark-gluon, quark-meson and quark-ghost interaction vertices}
\label{app:quarkvertices}

Here we discuss the constituents of our truncation that have not
been included in the previous works on quenched QCD \cite{Mitter:2014wpa} and 
Yang-Mills theory \cite{Cyrol:2016tym}.

\subsection{Two-quark-n-gluon vertices}
We classify the full tensor decomposition of the quark-gluon vertex,
the two-quark-two-gluon and the two-quark-three-gluon vertex in
tensors that can be related to operators of the type
$\bar q \slashed{D}^n q\,$, where
$D_\mu=\partial_\mu-\imag\, g\, \colT{a}\, A^a_\mu\,$. Therefore, the
elements of the full basis of each of the two-quark-$n$-gluon
interaction are ordered according to the number of explicit momentum
variables in the basis elements. Additionally, this expansion leads to
a natural classification in terms of chirally symmetric and
symmetry-breaking tensors. All even $n$ lead to operators that violate
chiral symmetry. In particular $n=0$ corresponds to the mass term in
the quark propagator.

\subsubsection{Quark-gluon vertex}

We expand the transversely projected quark-gluon vertex as in
\cite{Mitter:2014wpa},
\begin{align}
	\label{eq:barqAq}
	\Pi^{\trans}_{\mu\nu}(p+q)&\left[\Gamma^{(3)}_{\bar q q A}\right]^a_{\nu}(p,q) =\identityflavour\,  \colT{a}\,\\
	&\times\sum\limits_{i=1}^{8}\lambda^{(i)}_{\bar q q A}(p,q) 
	\Pi^\trans_{\mu\nu}(p+q)\left[{\cal T}^{(i)}_{\bar q q A}\right]_\nu(p,q)\,.\nonumber
\end{align}
For quantitative accuracy, we include the full three-dimensional
momentum dependence of the quark-gluon vertex dressing function. The
vertex has been assumed diagonal in flavour space via
$\identityflavour$ and $\colT{a}$ are the generators of the fundamental
representation of the $SU(3)$ colour group. The tensors
$\left[{\cal T}^{(i)}_{\bar q q A}\right]$ are given by
\begin{align}
\label{eq:barqAq_tensors}
  \bar q \slashed{D} q:\quad & \left[{\cal T}^{(1)}_{\bar q q A}\right]_\mu(p,q)=
                               -\imag\, \gamma_\mu\ ,\eqnewline
  \bar q \slashed{D}^2 q:\quad & \left[{\cal T}^{(2)}_{\bar q q A}\right]_\mu(p,q)=
                                 (p-q)_\mu\, \identityspinor\ ,\eqnewline
                             & \left[{\cal T}^{(3)}_{\bar q q A}\right]_\mu(p,q)=
                               (\slashed p-\slashed q)\gamma_\mu\ ,\eqnewline
                             & \left[{\cal T}^{(4)}_{\bar q q A}\right]_\mu(p,q)=
                               (\slashed p+\slashed q)\gamma_\mu\ ,\eqnewline
  \bar q \slashed{D}^3 q:\quad & \left[{\cal T}^{(5)}_{\bar q q A}\right]_\mu(p,q)=
                                 \imag\, (\slashed p+\slashed q) (p-q)_\mu\ ,\eqnewline
                             & \left[{\cal T}^{(6)}_{\bar q q A}\right]_\mu(p,q)=
                               \imag\, (\slashed p-\slashed q) (p-q)_\mu\ ,\eqnewline
                             & \left[{\cal T}^{(7)}_{\bar q q A}\right]_\mu(p,q)=
                               \tfrac{\imag}{2}[\slashed p,\slashed q]\gamma_\mu\ ,\eqnewline
  \bar q \slashed{D}^4 q:\quad & \left[{\cal T}^{(8)}_{\bar q q A}\right]_\mu(p,q)=
                                 -\tfrac{1}{2}[\slashed p,\slashed q](p-q)_\mu\ ,
\end{align}
and the transverse and longitudinal projection operators are given by
\begin{align}\label{eq:projectors}
  \Pi^\trans_{\mu\nu}(k)=&\, \delta_{\mu\nu}-\0{k_\mu k_\nu}{k^2}\,,
	\qquad \Pi^{\lt}_{\mu\nu}(k) = \0{k_\mu k_\nu}{k^2}\,.
\end{align}
The transverse projection of the tensors \eq{eq:barqAq_tensors} yields a basis for the transversely projected quark-gluon
vertex \eq{eq:barqAq}. Furthermore, the longitudinal projections of these tensors span the longitudinally projected vertex. However, only
four of the longitudinally projected tensors \eq{eq:barqAq_tensors} are linearly independent. We choose
\begin{align}\label{eq:barqAqlong_tensors}
&\left[{\cal T}^{(9)}_{\bar q q A}\right]_\mu(p,q)=\Pi^\lt_{\mu\nu}(p+q)\left[{\cal T}^{(1)}_{\bar q q A}\right]_\nu(p,q)\ ,\eqnewline
&\left[{\cal T}^{(10)}_{\bar q q A}\right]_\mu(p,q)=\Pi^\lt_{\mu\nu}(p+q)\left[{\cal T}^{(2)}_{\bar q q A}\right]_\nu(p,q)\ ,\eqnewline
&\left[{\cal T}^{(11)}_{\bar q q A}\right]_\mu(p,q)=\Pi^\lt_{\mu\nu}(p+q)\left[{\cal T}^{(6)}_{\bar q q A}\right]_\nu(p,q)\ ,\eqnewline
&\left[{\cal T}^{(12)}_{\bar q q A}\right]_\mu(p,q)=\Pi^\lt_{\mu\nu}(p+q)\left[{\cal T}^{(8)}_{\bar q q A}\right]_\nu(p,q)\ ,
\end{align}
as basis elements for the longitudinally projected quark-gluon vertex
\begin{align}\label{eq:barqAq_L}
  \Pi^\lt_{\mu\nu}(p+q)&\left[\Gamma^{(3)}_{\bar q q A}\right]^a_{\nu}(p,q) = \identityflavour\,  \colT{a}\, \nonumber\\
&\times\sum\limits_{i=9}^{12}\lambda^{(i)}_{\bar q q A}(p,q)\left[{
\cal T}^{(i)}_{\bar q q A}\right]_\mu(p,q)\,.
\end{align}
The full quark-gluon vertex is then given as the sum of \eq{eq:barqAq} and \eq{eq:barqAq_L}.

At this point, we want to discuss the consequences of a generalised
regularity assumption, which implies that the full quark-gluon vertex is 
spanned by the unprojected tensors \eq{eq:barqAq_tensors}, \ie
\begin{align}
	\label{eq:barqAq_reg}
	\left[\Gamma^{(3)}_{\bar q q A}\right]^a_{\nu}(p,q) &
		\stackrel{!}{=}\identityflavour\,  \colT{a}\,\nonumber\\
	&\times\sum\limits_{i=1}^{8}\lambda^{(i)}_{\bar q q A}(p,q)
		\left[{\cal T}^{(i)}_{\bar q q A}\right]_\nu(p,q)\,.
\end{align}
In the limit of vanishing gluon momentum, a regular 
quark-gluon vertex can be expressed in this way since 
singularities would otherwise be introduced by 
the projectors \eq{eq:projectors}.
Assuming that \eq{eq:barqAq_reg} holds also for finite gluon momenta, \ie 
assuming generalised regularity, the transverse and 
longitudinal dressing functions are obviously not independent any more.
By construction the transverse dressing functions of \eq{eq:barqAq} are 
identical to the dressings of \eq{eq:barqAq_reg}. On the other hand, the 
longitudinal dressings in \eq{eq:barqAq_L} are then given by particular
linear combinations of the transverse dressings
\begin{align}
\label{eq:longregassumption}
 \lambda^{(9)}_{\bar q q A}&=\left[\lambda^{(1)}_{\bar q q A}+\left(\frac{\lambda^{(7)}_{\bar q q A}}{2}-\lambda^{(5)}_{\bar q q A}\right)\left(p^2-q^2\right)\right]\,,\eqnewline
 \lambda^{(10)}_{\bar q q A}&=\left[\lambda^{(2)}_{\bar q q A}+\lambda^{(3)}_{\bar q q A}+\lambda^{(4)}_{\bar q q A}\frac{(p+q)^2}{p^2-q^2}\right]\,,\eqnewline
  \lambda^{(11)}_{\bar q q A}&=\left[\lambda^{(6)}_{\bar q q A}+\frac{1}{2}\lambda^{(7)}_{\bar q q A}\frac{(p+q)^2}{q^2-p^2}\right]\,,\eqnewline
   \lambda^{(12)}_{\bar q q A}&=\left[\lambda^{(8)}_{\bar q q A}+2\lambda^{(3)}_{\bar q q A}\frac{1}{q^2-p^2}\right]\,.
\end{align}
Here, we have abbreviated 
$\lambda^{(i)}_{\bar q q A}\equiv\lambda^{(i)}_{\bar q q A}(p,q)\,$. 
The four longitudinal dressings are constrained by the quark-gluon vertex 
STI, see \App{app:quarkghostscatteringkernel}.
Therefore, even under the generalized regularity assumption 
\eq{eq:barqAq_reg}, the STI constrains only the combinations 
of the transverse dressing functions given by the right hand sides
of \eq{eq:longregassumption}.

\subsubsection{Two-quark-two-gluon vertex}
The transversely projected two-quark-two-gluon vertex (projection operators suppressed for readability)
\begin{align}\label{eq:barqAAq}
  &\left[\Gamma^{(4)}_{\bar q q A^2}\right]_{\mu\nu}^{ab}(p_1,p_2,p_3) \\&\quad= \identityflavour\, \sum\limits_{i=1}^{18}\lambda^{(i)}_{\bar q q A^2}(\psym)\left[{\cal T}^{(i)}_{\bar q q A^2}\right]_{\mu\nu}^{ab}(p_1,p_2,p_3,p_4)\ ,\nonumber
\end{align}
receives contributions from the operators $\bar q \slashed{D}^n q$ with 
$n\geq 2$. We take into account the tensors corresponding
to $n=2\,$, 
\begin{align}\label{eq:barqAAq_tensors_D2}
\bar q \slashed{D}^2 q:\quad & 
\left[{\cal T}^{(16)}_{\bar q q 
A^2}\right]_{\mu\nu}^{ab}=\gamma_\mu\gamma_\nu \colT{a} \colT{b} + 
\gamma_\nu\gamma_\mu \colT{b} \colT{a}\ ,\eqnewline
& \left[{\cal T}^{(17)}_{\bar q q A^2}\right]_{\mu\nu}^{ab}=\gamma_\mu\gamma_\nu \colT{b} \colT{a} + \gamma_\nu\gamma_\mu \colT{a} \colT{b}\ ,\eqnewline
& \left[{\cal T}^{(18)}_{\bar q q A^2}\right]_{\mu\nu}^{ab}=
	\left(\gamma_\mu\gamma_\nu  + \gamma_\nu\gamma_\mu\right)\delta^{ab}\ ,
\end{align}
and $n=3\,$,
\begin{align}\label{eq:barqAAq_tensors_D3}
\bar q \slashed{D}^3 q:\quad & \left[{\cal T}^{(1)}_{\bar q q A^2}\right]_{\mu\nu}^{ab}=\imag \left(p_{1,\mu}\gamma_\nu+p_{1,\nu}\gamma_\mu\right)\delta^{ab}\ ,\\[2ex]
& \left[{\cal T}^{(2)}_{\bar q q A^2}\right]_{\mu\nu}^{ab}=\imag \left(p_{2,\mu}\gamma_\nu+p_{2,\nu}\gamma_\mu\right)\delta^{ab}\ ,\eqnewline
& \left[{\cal T}^{(3)}_{\bar q q A^2}\right]_{\mu\nu}^{ab}=\imag \left(p_{1,\mu}\gamma_\nu \colT{a}\colT{b}+p_{1,\nu}\gamma_\mu \colT{b}\colT{a}\right)\ ,\eqnewline
& \left[{\cal T}^{(4)}_{\bar q q A^2}\right]_{\mu\nu}^{ab}=\imag \left(p_{2,\mu}\gamma_\nu \colT{a}\colT{b}+p_{2,\nu}\gamma_\mu \colT{b}\colT{a}\right)\ ,\eqnewline
& \left[{\cal T}^{(5)}_{\bar q q A^2}\right]_{\mu\nu}^{ab}=\imag \left(p_{1,\mu}\gamma_\nu \colT{b}\colT{a}+p_{1,\nu}\gamma_\mu \colT{a}\colT{b}\right)\ ,\eqnewline
& \left[{\cal T}^{(6)}_{\bar q q A^2}\right]_{\mu\nu}^{ab}=\imag \left(p_{2,\mu}\gamma_\nu \colT{b}\colT{a}+p_{2,\nu}\gamma_\mu \colT{a}\colT{b}\right)\ ,\eqnewline
& \left[{\cal T}^{(7)}_{\bar q q A^2}\right]_{\mu\nu}^{ab}=\imag \left(\gamma_\mu\slashed{p_1}\gamma_\nu+\gamma_\nu\slashed{p_1}\gamma_\mu\right)\delta^{ab}\ ,\eqnewline
& \left[{\cal T}^{(8)}_{\bar q q A^2}\right]_{\mu\nu}^{ab}=\imag \left(\gamma_\mu\slashed{p_2}\gamma_\nu+\gamma_\nu\slashed{p_2}\gamma_\mu\right)\delta^{ab}\ ,\eqnewline
& \left[{\cal T}^{(9)}_{\bar q q A^2}\right]_{\mu\nu}^{ab}=\imag \left(\gamma_\mu\slashed{p_1}\gamma_\nu \colT{a}\colT{b}+\gamma_\nu\slashed{p_1}\gamma_\mu \colT{b}\colT{a}\right)\ ,\eqnewline
& \left[{\cal T}^{(10)}_{\bar q q A^2}\right]_{\mu\nu}^{ab}=\imag \left(\gamma_\mu\slashed{p_2}\gamma_\nu \colT{a}\colT{b}+\gamma_\nu\slashed{p_2}\gamma_\mu \colT{b}\colT{a}\right)\ ,\eqnewline
& \left[{\cal T}^{(11)}_{\bar q q A^2}\right]_{\mu\nu}^{ab}=\imag \left(\gamma_\mu\slashed{p_1}\gamma_\nu \colT{b}\colT{a}+\gamma_\nu\slashed{p_1}\gamma_\mu \colT{a}\colT{b}\right)\ ,\eqnewline
& \left[{\cal T}^{(12)}_{\bar q q A^2}\right]_{\mu\nu}^{ab}=\imag \left(\gamma_\mu\slashed{p_2}\gamma_\nu \colT{b}\colT{a}+\gamma_\nu\slashed{p_2}\gamma_\mu \colT{a}\colT{b}\right)\ ,\eqnewline
& \left[{\cal T}^{(13)}_{\bar q q A^2}\right]_{\mu\nu}^{ab}=\imag \left(\gamma_\mu\slashed{p_3}\gamma_\nu+\gamma_\nu\slashed{p_4}\gamma_\mu\right)\delta^{ab}\ ,\eqnewline
& \left[{\cal T}^{(14)}_{\bar q q A^2}\right]_{\mu\nu}^{ab}=\imag \left(\gamma_\mu\slashed{p_3}\gamma_\nu \colT{a}\colT{b}+\gamma_\nu\slashed{p_4}\gamma_\mu \colT{b}\colT{a}\right)\ ,\eqnewline
& \left[{\cal T}^{(15)}_{\bar q q A^2}\right]_{\mu\nu}^{ab}=\imag \left(\gamma_\mu\slashed{p_3}\gamma_\nu \colT{b}\colT{a}+\gamma_\nu\slashed{p_4}\gamma_\mu \colT{a}\colT{b}\right)\ ,\nonumber
\end{align}
that are symmetric under $A^{\mu,a}\leftrightarrow A^{\nu,b}\,$. 
We find that after transverse projection, all other conceivable tensors of 
this type can be expressed in terms of the tensors listed in 
\eq{eq:barqAAq_tensors_D2} and \eq{eq:barqAAq_tensors_D3}.

\subsubsection{Two-quark-three-gluon vertex}
The transversely projected two-quark-three-gluon vertex (projection operators suppressed for readability)
\begin{align}\label{eq:barqAAAq}
  &\left[\Gamma^{(5)}_{\bar q q A^3}\right]_{\mu\nu\rho}^{abc}(p_1,p_2,p_3,p_4) \\&\quad= \identityflavour\, \sum\limits_{i=1}^{5}\lambda^{(i)}_{\bar q q A^3}(\psym)\left[{\cal T}^{(i)}_{\bar q q A^3}\right]_{\mu\nu\rho}^{abc}(p_1,p_2,p_3,p_4,p_5)\ ,\nonumber
\end{align}
receives contributions from the operator $\slashed{D}^n q$ with $n\geq 3$. We take into account the tensors corresponding
to $n=3\,$, 
\begin{align}\label{eq:barqAAAq_tensors_D3}
\bar q \slashed{D}^3 q:\quad & \left[{\cal T}^{(1)}_{\bar q q A^3}\right]_{\mu\nu\rho}^{abc}=-\left(\gamma^\mu\gamma^\nu\gamma^\rho-\gamma^\mu\gamma^\rho\gamma^\nu+\dots\right)f^{abc}\ ,\eqnewline
& \left[{\cal T}^{(2)}_{\bar q q A^3}\right]_{\mu\nu\rho}^{abc}=\imag\left(\gamma^\mu\gamma^\nu\gamma^\rho \colT{a}\colT{b}\colT{c}+\text{ perm.}\right)\ ,\eqnewline
& \left[{\cal T}^{(3)}_{\bar q q A^3}\right]_{\mu\nu\rho}^{abc}=\imag\left(\gamma^\mu\gamma^\nu\gamma^\rho \colT{c}\colT{a}\colT{b}+\text{ perm.}\right)\ ,\eqnewline
& \left[{\cal T}^{(4)}_{\bar q q A^3}\right]_{\mu\nu\rho}^{abc}=\imag\left(\gamma^\mu\gamma^\nu\gamma^\rho \colT{b}\colT{c}\colT{a}+\text{ perm.}\right)\ ,\eqnewline
& \left[{\cal T}^{(5)}_{\bar q q A^3}\right]_{\mu\nu\rho}^{abc}=\imag\left(\gamma^\mu\gamma^\nu\gamma^\rho \colT{b}\colT{a}\colT{c}+\text{ perm.}\right)\ ,
\end{align}
that are symmetric under $A^{\mu,a}\leftrightarrow 
A^{\nu,b}\leftrightarrow A^{\rho,c}$. 

\subsection{Two-quark-n-meson vertices}
\label{app:twoquarknmeson}
\subsubsection{Yukawa interaction vertex}
We expand the quark-pion Yukawa interaction as \cite{LlewellynSmith:1969az}
\begin{align}\label{eq:barqpiq}
  \left[\Gamma^{(3)}_{\bar q q \pi}\right]^a(p,q) = \identitycolour\, &\flavT{a}\, \imag\, \gamma_5\, \\\times&\sum\limits_{i=1}^{4}h^{(i)}_{\bar q q \pi}(p_{\rm soft})\,  \left[{\cal T}^{(i)}_{\bar q q \pi}\right](p,q)\ ,\nonumber
\end{align}
with $a \in\{1,\, 2,\, 3\}$ and tensor basis elements
\begin{align}\label{eq:barqpiq_tensors}
 \left[{\cal T}^{(1)}_{\bar q q \pi}\right](p,q)&=\identityspinor\ ,\eqnewline
 \left[{\cal T}^{(2)}_{\bar q q \pi}\right](p,q)&=\imag\left(\slashed{p} + \slashed{q} \right)\ ,\eqnewline
 \left[{\cal T}^{(3)}_{\bar q q \pi}\right](p,q)&=\imag\left(\slashed{p} - \slashed{q} \right)\ ,\eqnewline
 \left[{\cal T}^{(4)}_{\bar q q \pi}\right](p,q)&=-\frac{1}{2}\, \left[\slashed{p},\slashed{q}\right]\ .
\end{align}
Here, $\flavT{a}\,$, $a\in\{1,2,3\}$ are the generators of the
$SU(2)$ flavour group with $\Tr (\flavT{a} \flavT{b})=\delta^{ab} /2\,$. We 
calculate the
momentum dependence of the dressing functions of the Yukawa
interaction from the soft-pion channel, where the quark and antiquark have
opposite momenta. This is the channel that is most important for the
momentum-dependent version of the dynamical hadronisation used in this
work, see \App{app:dyn_had}. The full momentum dependence is
approximated from this single-momentum channel via
\begin{align}
	h^{(i)}_{\bar q q \pi}(p_1,p_2) &= 
	h^{(i)}_{\bar q q\pi}
		\left(\sqrt{\frac{(p_1-p_2)^2}{4}+(p_1+p_2)^2}\right)\,.
\end{align}
This choice is the outcome of explicit checks of the momentum
dependence of the Yukawa interaction. We find that the interaction
strength drops faster by a factor of two with the pion momentum
$p_1+p_2\,$, which is reflected by the above formula.

\subsubsection{Higher order quark-meson interactions}
Furthermore, we add the simplest tensor structures that contribute to
the two-quark-two-pion and the two-quark-three-pion vertex. These
tensor structures are created from field dependencies in the
Yukawa interaction $h^{(1)}_{\bar q q \pi}(\rho)$ and have been found to 
yield small corrections in \cite{Pawlowski:2014zaa}. The vertices are then 
given by
\begin{align}
	\label{eq:barqpinq}
	\left[\Gamma^{(4)}_{\bar q q \pi^2}\right]^{ab}(\{p_i\}) &= 
	h^{(1)}_{\bar q q \pi^2}(p_{\rm soft})\,
		\left[{\cal T}_{\bar q q \pi^2}\right]^{ab}\,,\eqnewline
	\left[\Gamma^{(5)}_{\bar q q \pi^3}\right]^{abc}(\{p_i\}) &= 
	h^{(1)}_{\bar q q \pi^3}(p_{\rm soft})\,
		\left[{\cal T}_{\bar q q \pi^3}\right]^{abc}\,,
\end{align}
with the tensor basis elements
\begin{align}\label{eq:barqpinq_tensors}
  \left[{\cal T}_{\bar q q \pi^2}\right]^{ab}&= 
	\identitycolour\, \identityspinor\, \flavT{0}\, \delta^{ab}\,,\eqnewline
  \left[{\cal T}_{\bar q q \pi^3}\right]^{abc}&=
	\imag \, \identitycolour\, \gamma_5\, 
\left(\delta^{ab}\flavT{c}+\delta^{bc}\flavT{a}+\delta^{ca}\flavT{b}\right)\,,
\end{align}
where $a,\, b,\, c\in \{1,\,2,\, 3\}\,$. Since these interactions stem from
field-dependencies in the Yukawa interaction $h^{(1)}_{\bar q q \pi}(\rho)$, they
have to be proportional to each other at vanishing external momenta
\begin{align}
	\label{eq:2q2m-relation}
	h^{(1)}_{\bar q q \pi^2}(0) &=
		h^{(1)}_{\bar q q \pi^3}(0)\langle \sigma\rangle\,,
\end{align}
which we find to be fulfilled to \SI{2}{\percent} in our numerical results.
Similarly to the Yukawa interaction, we approximate the full momentum
dependence of the two-quark-$n$-pion interactions from the soft-pion
momentum channel via
\begin{align}
	h^{(1)}_{\bar q q \pi^n}(\{p_i\}) &= 
		h^{(1)}_{\bar q q \pi^n}
		\left(\sqrt{\frac{(p_1-p_2)^2}{4}+p_3^2+\dots+p_{n+2}^2}\right)\,.
\end{align}
for $n\in\{2,3\}\,$.

Due to chiral symmetry, the two-quark-$n$-sigma interactions can be
obtained from combinations of the corresponding quark-pion
interactions and are given in our truncation by
\begin{align}\label{eq:barqsigmaq}
  \left[\Gamma^{(3)}_{\bar q q \sigma}\right](p,q) &= \identitycolour\, \flavT{0}\, \left[{\cal T}^{(1)}_{\bar q q \pi}\right](p,q)\nonumber\\
	&\times \left(h^{(1)}_{\bar q q \pi}(p_{\rm soft})+\langle \sigma\rangle^2 h^{(1)}_{\bar q q \phi^3}(p_{\rm soft}) \right)\ ,\nonumber\\
  \left[\Gamma^{(4)}_{\bar q q \sigma^2}\right](\{p_i\}) &= 3\, \langle\sigma\rangle\, h^{(1)}_{\bar q q \pi^3}(p_{\rm soft})\,  \left[{\cal T}_{\bar q q \pi^2}\right]^{00}\ ,\nonumber\\
  \left[\Gamma^{(5)}_{\bar q q \sigma^3}\right](\{p_i\}) &= 3\, h^{(1)}_{\bar q q \pi^3}(p_{\rm soft})\,  \left[{\cal T}_{\bar q q \pi^2}\right]^{00}\ .
\end{align}
Here we took only the first tensor structure for the quark-sigma
Yukawa interaction into account, since the sigma-meson effects are
considerably suppressed in comparison to the pion
effects. Furthermore, the two-quark-two-sigma interaction has been
approximated by using the relation \eq{eq:2q2m-relation} and
contributions that would come from higher field dependencies in
$h^{(1)}_{\bar q q \pi}(\rho)$ have not been taken into account. On
the one hand, they are not present in our current truncation, and, on
the other hand, their effect has been found to be negligible
\cite{Pawlowski:2014zaa}.

\section{Modified STI and quark-ghost scattering kernel}
\label{app:quarkghostscatteringkernel}

The quark-ghost scattering kernel plays a crucial r\^ole in the
Slavnov-Taylor identity of the quark-gluon vertex. For the sake of
completeness we briefly sketch the derivation of the latter in the
presence of infrared cutoff terms, 
see \cite{Ellwanger:1994iz,D'Attanasio:1996jd,Igarashi:2001mf,Pawlowski:2005xe}. 
The related master equation is derived from the generating functional 
$Z_k[J,Q]$ in the presence of source terms for the BRST transformations,
\begin{align}\label{eq:ZBRST} 
  Z_k[J,Q]=\int \dd\Phi \dd\lambda\, e^{-S_k[\Phi]+\Delta S_k[\Phi]+
\int_x  J_i \varphi_i+\int_x  Q_i\, \sBRST \varphi_i}\,, 
\end{align}
with the currents $Q_{A}, Q_{c}, Q_{{\bar c}}, Q_{q}, Q_{{\bar q}}\,$,
or short $Q_i$ for $Q_{\varphi_i}$, for the BRST transformations of
the fundamental fields,
\begin{align}\label{eq:sBRST} \sBRST (A_\mu, c, \bar c, q,\bar q) =\epsilon 
  ( D_\mu c\,,\, \imag g\,  c^2\, ,\, \lambda\,,\, 
\imag g \, c q\,,\,  \imag g \,\bar q c )\,, 
\end{align}
with a Grassmann number $\epsilon\,$. The BRST transformations as well
as the BRST currents of the fermionic fields are Grassmannian, while
those of the bosonic fields are $c$-numbers. This renders the source
terms $c$-numbers. The BRST transformation of $\lambda$ is given by
$\sBRST \lambda=0\,$. We also have $\sBRST \phi=0$ for the colourless
mesonic fields. Consequently, we set
$Q_\lambda=Q_\phi=0\,$.  The effective action
\begin{align}
\label{eq:Glen}
\Gamma_k[\Phi,Q]= -W_k[J,Q] +\int_x J_i \varphi_i-\Delta S_k[\Phi]\,
\end{align} 
is the Legendre transform of $W_k[J,Q]=\log Z_k[J,Q]$ w.r.t.\
$J\,$, and hence has the same $Q$-derivatives as the Schwinger
functional $W_k\,$. This entails that the generator of quantum BRST
transformations commutes with all fields. It reads in terms of the
effective action,
\begin{align}\label{eq:genBRST} 
  -\0{\delta \Gamma_k}{\delta Q_i}\0{\delta}{\delta \varphi_i} \,,\qquad 
  {\rm with} \quad  -\0{\delta \Gamma_k}{\delta Q_\Phi} = 
  \0{\delta W_k}{\delta Q_\Phi}=\langle 
  \sBRST \Phi\rangle\,.
\end{align}
The gauge-fixed action is invariant under 
BRST transformations, as is the BRST source term with
$\sBRST^2=0\,$. The invariance of the path integral measure in
\eq{eq:ZBRST} under BRST transformation then leads to the modified
master equation,
\begin{align}\label{eq:master} 
\int_x  \0{\delta \Gamma_k}{\delta Q_i}\,\0{\delta \Gamma_k}{\delta \varphi_i}  
  = \Tr \, (R_k\, G_k)_{i j} \, \0{\delta^2 \Gamma_k}{\delta \varphi_j  \delta Q_i}\,, 
\end{align}
with the full field-dependent propagator $G_k$ defined in
\eq{eq:Gn}. The second term in \eq{eq:master} vanishes at
$R_k\equiv 0\,$, leaving us with the standard homogeneous master equation
at $k=0\,$. The modified STIs (mSTIs) for correlation functions are 
obtained
from the master equation \eq{eq:master} by taking appropriate derivatives 
w.r.t.\ to the fields.

The generating equation for the mSTIs \eq{eq:master} depends on the BRST variations of the
effective action. They are (generalised) vertices of the theory and
can be straightforwardly computed from \eq{eq:sBRST} and
\eq{eq:genBRST}. We find 
\begin{align}
	\Gamma_{Q^{\!}_{\!\Phi}}^{(1)} =
	&\,-\epsilon  \Bigl( D_\mu c+G_{Ac}\,,
		\,\imag g\left(c^2+G_{cc}\right)\, ,\, \lambda\eqnewline
	& \hspace{.5cm}\,,\,\imag g \left( c q+G_{cq}\right)\,,
		\,\imag g \left( \bar q c +G_{\bar q c}\right)\Bigr)\,,  
	\label{eq:sQuantum}
\end{align}
with
$G_{\varphi_i\varphi_j} = \left( G_k \right)_{\varphi_i\varphi_j}$ being
the $\varphi_i,\varphi_j$ component of the full propagator. Note that we 
have used in an abuse of notation the same symbol for the fields and their 
expectation values as they appear as argument in the effective action.
The quantum BRST transformations are therefore obtained from the 
classical BRST transformations by the replacement
\begin{align}
\varphi_i(x) \varphi_j(x) \to \varphi_i(x) \varphi_j(x) + G_{\varphi_i \varphi_j}(x,x)\,, 
\end{align}
while the linear part is unchanged. Accordingly, the quantum
deformation consists of the diagonal part of the mixed propagator
$G_{\varphi_i \varphi_j}(x,x)\,$ and is given by a loop in momentum space. 
For the BRST
transformation of the gauge field,
\begin{align}\label{eq:sQuantumA}
\Gamma^{(1)}_{Q^{\ }_{\! A}}=-\epsilon\langle D_\mu c\rangle \,,
\end{align} 
we can invoke the anti-ghost Dyson-Schwinger equation (DSE). To that
end we notice that the anti-ghost DSE, derived via a derivative w.r.t.\ $\partial_\mu \bar c$,
can be written as
\begin{align}\label{eq:barcDSE}
\0{1}{\partial^2}\partial_\mu \0{\delta\Gamma_k}{ \delta \bar c} = 
\langle D_\mu \,c\rangle\,.
\end{align}
Inserting \eq{eq:barcDSE} in \eq{eq:sQuantumA} leaves us with
\begin{align}\label{eq:sQuantumAfinal}
\Gamma^{(1)}_{Q^{\ }_{\!A}}=-\epsilon 
\partial_\mu  \0{1}{\partial^2} \Gamma^{(1)}_{\bar c}{}^{a} \,, 
\end{align} 
as an alternative representation for the first term in
\eq{eq:sQuantum}. This is preferable, as it depends on a first derivative
of the effective action instead of the inverse of the full two-point function.

The STI for the quark-gluon vertex is now derived from \eq{eq:master}
by taking a $q\,$, $\bar q$ and $c$-derivative at vanishing fields, 
\begin{align}\label{eq:STIqGproject} 
  \left. \0{\delta}{\delta q} \0{\delta}{\delta \bar q} \0{\delta}{\delta c} 
\eq{eq:master}\right|_{\Phi,\lambda=0} \,. 
\end{align}
Due to the ghost derivative, only terms with ghost number one are
left. On both sides of \eq{eq:master} these are the terms that involve
BRST variations of the gluon, quark and anti-quark fields. For the rest of the
current work we drop the right hand side of \eq{eq:master} also at
finite $k$, leading to
\begin{align}\label{eq:STIqg}
  \int_x\left(  \Gamma^{(2)}_{Q_A c}\Gamma^{(3)}_{A \bar q q }-  
 \Gamma^{(2)}_{\bar q q } \Gamma^{(3)}_{Q_q c  q}+ 
  \Gamma^{(3)}_{Q_{ \bar q} c \bar q} \Gamma^{(2)}_{\bar q q}\right) = 0\,. 
\end{align} 
This approximation guarantees the standard STI at $k=0\,$. The
modification terms on the right hand side of the master equation at
finite $k$ only feed back as sub-leading terms in other flow equations
beyond the accuracy of the current approximation, see also the
discussion below \eq{eq:flowBRST}. This will be resolved in a future
work where a mSTI-consistent approximation will be used, thus
resolving the current issues.

It is left to compute the unknown vertex functions
$\Gamma^{(2)}_{c Q_A}$, $\Gamma^{(3)}_{qc Q_q}\,$,
$\Gamma^{(3)}_{c \bar q Q_{\bar q}}$ in \eq{eq:STIqg} from
$\Gamma^{(1)}_{Q_\Phi}$ in \eq{eq:sQuantum} and \eq{eq:sQuantumA}. As
it is simply given by a momentum-space loop of one propagator, the
quantum deformation in \eq{eq:sQuantum} requires in general a
regularisation.  Within the present FRG approach this is consistently
resolved with the flow equations for the quantum BRST
transformations. These are most easily obtained by taking a
$Q_l$-derivative of the flow of $\Gamma_k\,$,
\begin{align}\label{eq:flowBRST}
  \partial_t \0{\delta \Gamma_k}{\delta Q_l } = -\012 \Tr \, ( G_k 
  \partial_t R_k G_k)_{ij} 
  \0{\delta^3  \Gamma_k}{\delta \varphi_j\delta \varphi_i \delta  Q_l} \,.
\end{align}
We shall solve this flow in the following
approximation: we rewrite its right hand side as a total derivative
w.r.t.~$t$ and terms proportional to $\partial_t \Gamma^{(n)}$, 
\begin{align}\label{eq:flowBRST2}
  \partial_t \0{\delta \Gamma_k}{\delta Q_l } = \partial_t \left( \012 \Tr \,  G_{k,ij} 
  \0{\delta^3  \Gamma_k}{\delta \varphi_j\delta \varphi_i 
  \delta  Q_l} \right) + \partial_t \Gamma^{(n)}{\rm -terms}\,.
\end{align}
The latter RG-improvement terms are dropped as they are higher order
in fully RG-resummed loops, and hence are higher order corrections of
the BRST variations. Note that this even reproduces the correct result
if only the classical BRST variation
$S_{\text{\tiny{QCD}},\varphi_j\varphi_i j Q_l}^{(3)}$ is inserted in
the right hand side. 

In this approximation, we get
\begin{align} \label{eq:GAQ-Gcbarc}
\Gamma^{(2)}_{Q_A c}{}^{ab}_\mu = -\epsilon 
\partial_\mu  \0{1}{\partial^2} \Gamma^{(2)}_{\bar c c}{}^{ab}\,.
\end{align}
The remaining two terms, $\Gamma^{(3)}_{Q_q c q}$ and
$\Gamma^{(3)}_{ Q_{\bar q} c \bar q}\,$, are similar and we only discuss
the former. With \eq{eq:sQuantum} we are led to
\begin{align} \label{eq:GqcQ}
\Gamma^{(3)}_{Q_q c q} = - \imag g \epsilon \0{\delta}{\delta 
q}\0{\delta}{\delta c^a} \left(  c\, q + G_{k,cq}\right)\,. 
\end{align}
We expand the quark-ghost scattering kernel \eq{eq:GqcQ} as
\begin{align}\label{eq:qcQq}
  \Gamma^{(3)}_{Q_q c q}  =-\imag \epsilon \sum_{i=1}^4  \lambda^{(i)}_{Q_q c q} \mathcal{T}^{(i)}_{Q_q c q}\,,
\end{align}
with four tensor structures $\mathcal{T}^{(i)}_{Q_q c q}\,$. In momentum
space it finally reads
\begin{align}\label{eq:qg-kernel}  \Gamma^{(3)}_{Q_q c q}(r,q,p) =& -\imag \epsilon\, \colT{a} \Big[
    \lambda^{(1)}_{Q_q c q}(r,p) +\lambda^{(2)}_{Q_q c q}(r,p)
    \slashed{r} \eqnewline & +\lambda^{(3)}_{Q_q c q}(r,p) \,
    \slashed{p} + \lambda^{(4)}_{Q_q c q}(r,p) \,
    \frac{1}{2}\left[\slashed{r},\slashed{p}\right]\Bigr]\,.
\end{align}
The above expansion coefficients relate to the expansion
  coefficients $X_{i}$ that are predominantly used in the literature
  \cite{Davydychev:2000rt,Aguilar:2010cn,Rojas:2013tza,Aguilar:2016lbe}
  via $\lambda^{(i)}_{Q_q c q}(p,q)=g X_{i-1}(p,q)$ for 
  $i\in\{1,3\}\,$ and $\lambda^{(i)}_{Q_q c q}(p,q)=-g X_{i-1}(p,q)$ for $i\in\{2,4\}\,$. 
  Equation \eq{eq:qg-kernel} involves loop terms
originating from the $cq$-propagator in \eq{eq:GqcQ} as well as a
tree-level part $\imag g$, originating from the expectation values in
\eq{eq:GqcQ}. The latter part reflects the classical BRST
transformations \eq{eq:sBRST} with
\begin{align}
	\label{eq:qg-classical}
	\left.\lambda^{(i)}_{Q_q c q}\right|_{\rm cl} =\delta_{i1} g \,.
\end{align}
In the current approximation, the quark-ghost scattering kernel fulfils 
the equation, see \eg \cite{Alkofer:2000wg},
\begin{align}
  \Gamma^{(3)}_{Q_q c q} = -\imag\, g\,\epsilon\, \colT{a} \ -
 \vcenter{\hbox{\includegraphics[width=0.15\textwidth]{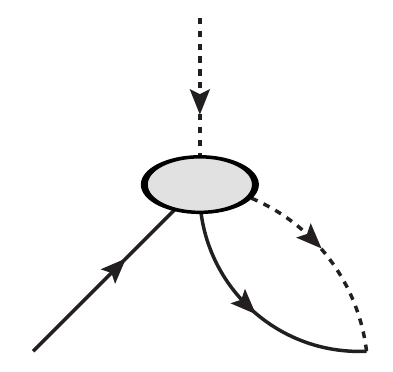}}}\ .
\end{align}
This involves the connected two-quark-two-ghost vertex as well as the
dressed quark and ghost propagators. Ignoring its sub-leading $1$PI
part, the two-quark-two-ghost vertex is obtained from the dressed
quark-gluon and ghost-gluon vertices, connected by a dressed gluon
propagator. In this approximation, the quark-ghost scattering kernel
takes the form, see \eg \cite{Aguilar:2010cn,Rojas:2013tza,Aguilar:2016lbe}, 
\begin{align}
  \Gamma^{(3)}_{Q_q c q} = -\imag\, g\, \epsilon\, \colT{a} \ -
  \vcenter{\hbox{\includegraphics[width=0.15\textwidth]{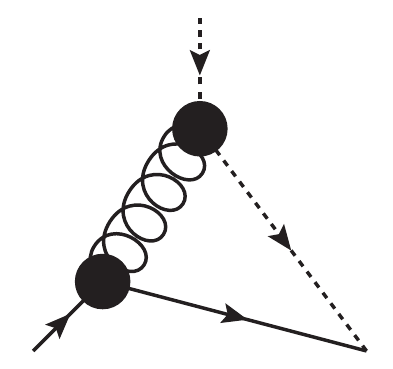}}}\ .
\end{align}
The dressing functions of the quark-ghost scattering kernel
\eq{eq:qg-kernel} are obtained by evaluating
appropriate projections of this simple integral. Finally, these dressing
functions are used in the Slavnov-Taylor identity \eq{eq:STI} for the
quark-gluon vertex. 

Note that $\Gamma^{(3)}_{c\bar q Q_{\bar q}}$ has an analogous expansion 
in terms of the same expansion functions as in \eq{eq:qg-kernel} 
\cite{Davydychev:2000rt,Aguilar:2010cn,Rojas:2013tza,Aguilar:2016lbe},
\begin{align}\label{eq:qbarg-kernel}  \Gamma^{(3)}_{Q_{\bar q}c\bar q }(r,q,p) =& -\imag \epsilon\, \colT{a} \Big[  
    \lambda^{(1)}_{Q_q c q}(p,r) -\lambda^{(2)}_{Q_q c q}(p,r)
    \slashed{p} \eqnewline & -\lambda^{(3)}_{Q_q c q}(p,r) \,
    \slashed{r} - \lambda^{(4)}_{Q_q c q}(r,p) \,
    \frac{1}{2}\left[\slashed{p},\slashed{r}\right]\Bigr]\,.
\end{align}
It remains to establish the connection between \eq{eq:STI} and \eq{eq:STIqg}. Note
therefore that \eq{eq:STIqg} reads in momentum space
\begin{align}
0=\int_q &\Gamma^{(2)}_{Q_A c}(-q,p)\Gamma^{(3)}_{ A \bar q q}(q,p_1,p_2)\eqnewline
&-\Gamma^{(2)}_{\bar q q}(p_1,-q)\Gamma^{(3)}_{Q_q c q}(q,p,p_2)\eqnewline
&+\Gamma^{(3)}_{ Q_{\bar q} c\bar q}(q,p,p_1)\Gamma^{(2)}_{\bar q q}(-q,p_2)
\end{align}
suppressing all external indices and denoting the ghost momentum by $p\,$ 
and the (anti-)quark momenta by $p_2$($p_1$). After inserting propagators, 
the relation reads
\begin{align}
0=&-\epsilon \colT{a} Z_c(p^2)(-\imag p_\mu) [\Gamma^{(3)}_{\bar q q A}(p_1,p_2,p)]^\mu \eqnewline
&-Z_q(p_1^2)(\imag \slashed p_1+M_q(p_1^2))\Gamma^{(3)}_{Q_q c q}(p_1,p,p_2)\eqnewline
&+\Gamma^{(3)}_{ Q_{\bar q} c\bar q}(p_2,p,p_1)Z_q(p_2^2)(\imag \slashed p_2+M_q(p_2^2))
\end{align}
Inserting the Ans\"atze \eq{eq:qg-kernel}, \eq{eq:qbarg-kernel} and \eq{eq:barqAq_L}, we obtain
\begin{align}
  0=& -\imag Z_c(p^2)(p_1+p_2)_{\mu} \sum_{i=9}^{12} \lambda^{(i)}_{\bar q q A}(p_1,p_2) \mathcal{T}^{(i)}_{\bar q q A}(p_1,p_2)^{\mu} \eqnewline
    &-Z_q(p_1^2)(\imag \slashed p_1+M_q(p_1^2))\Gamma^{(3)}_{Q_q c q}(p_1,-p_1-p_2,p_2)\eqnewline
    &+\Gamma^{(3)}_{ Q_{\bar q} c\bar q}(p_2,-p_1-p_2,p_1) Z_q(p_2^2)(\imag 
\slashed p_2+M_q(p_2^2))\,,
\end{align}
which still carries a Dirac structure. We can
contract the equation with either $\identityspinor\,$,
$\slashed p_1+\slashed p_2$, $\slashed p_1-\slashed p_2$ or
$\tfrac{1}{2}[\slashed p_1,\slashed p_2]$ to obtain explicit
STIs for $\lambda^{(i)}_{\bar q q A}$ for
$i\in\{9,\ldots,12\}\,$.
This leads to 
\begin{widetext}
\begin{align}
\lambda^{(9)}_{\bar q q A}(p_1,p_2)=&\frac{ 1 }{2Z_c(p)}\Biggl( Z_q(p_2) \left(\lambda^{(1)}_{12} - \imag \left(\lambda^{(2)}_{12}+\lambda^{(3)}_{12}\right)M_q(p_2)+ \lambda^{(4)}_{12} (p_1-p_2)\cdot p_2\right)\eqnewline
&+Z_q(p_1) \left(\lambda^{(1)}_{21} + \imag \left(\lambda^{(2)}_{21}+\lambda^{(3)}_{21}\right)M_q(p_1)+ \lambda^{(4)}_{21} (p_2-p_1)\cdot p_1\right)\Biggr)\,,\eqnewline
\lambda^{(10)}_{\bar q q A}(p_1,p_2)=&\frac{ 1 }{2Z_c(p)(p_1^2-p_2^2)}\Biggl( Z_q(p_2) \left(-\lambda^{(1)}_{12}M_q(p_2)-\imag\lambda^{(2)}_{12}p_1^2 -\imag \lambda^{(3)}_{12}p_2^2\right)\eqnewline
&+Z_q(p_1) \left(-\lambda^{(1)}_{21} M_q(p_1)+\imag\lambda^{(2)}_{21}p_2^2 +\imag \lambda^{(3)}_{21}p_1^2\right)\Biggr)\,,\eqnewline
\lambda^{(11)}_{\bar q q A}(p_1,p_2)=&\frac{ 1 }{Z_c(p)(p_1^2-p_2^2)}\Biggl(\nonumber Z_q(p_2) \left(-\lambda^{(1)}_{12} - \imag \left(\lambda^{(2)}_{12} - \lambda^{(3)}_{12}\right) M_q(p_2) -\lambda^{(4)}_{12} (p_1+p_2)\cdot p_2)\right)\eqnewline
&+Z_q(p_1) \left(\lambda^{(1)}_{21} - \imag \left(\lambda^{(2)}_{21} - \lambda^{(3)}_{21}\right) M_q(p_1) +\lambda^{(4)}_{21} (p_1+p_2)\cdot p_1\right)\Biggr)\,,\eqnewline
\lambda^{(12)}_{\bar q q A}(p_1,p_2)=&\frac{ 1 }{Z_c(p)(p_1^2-p_2^2)}\Biggl(Z_q(p_2)\left(-\imag \lambda^{(2)}_{12} +\lambda^{(4)}_{12}M_q(p_2)\right) +Z_q(p_1)\left(\imag \lambda^{(2)}_{21} + \lambda^{(4)}_{21}M_q(p_1)\right)\Biggr)\,,
\end{align}
\end{widetext}
using the shorthand notation 
$\lambda^{(i)}_{ij}\equiv \lambda^{(i)}_{Q_qcq}(p_i,p_j)\,$. Evaluated at 
the symmetric point one obtains for $\lambda^{(9)}_{\bar q q A}(\psym)\,$:
\begin{align}
\lambda^{(9)}_{\bar q q A}(\psym)=
\frac{Z_q(\psym)}{Z_c(\psym)}\left(\lambda^{(1)}_{Q_qcq}(\psym)-\frac{3}{2} 
\lambda^{(4)}_{Q_qcq}(\psym) \psym^2\right)\,.
\end{align}

\section{Transverse Ward-Takahashi identities}
\label{app:tWTI}

Transverse Ward-Takahashi identities (tWTI)
\cite{Takahashi:1985yz,Kondo:1996xn,He:2000we,Qin:2013mta,Aguilar:2014lha} constitute
combinations of DSEs.  They are functional relations for exterior
derivatives $\langle D_\mu J_\nu - D_\nu J_\mu\rangle$ with fermionic
currents $J_\mu$, whereas the (longitudinal) WTIs or STIs are
relations for expectation values of covariantly conserved current,
$\langle D_\mu J_\mu\rangle\,$.

The derivation of the transverse WTIs is
based on the observation that DSEs for mixed gluon-quark correlation
functions can be derived from either the gluon DSE or the quark DSE, for a lucid discussion 
see \cite{Kondo:1996xn}. The formal equivalence of both DSEs leads to the
tWTI. An important difference to the longitudinal WTI or STI are contributions of the form
\begin{align}\label{eq:nonsym}  
  \left \langle D_\mu \0{\delta (S_{\text{\tiny YM}}+S_{\text{\tiny{ghost}}})} {\delta A_\nu}- 
D_\nu \0{\delta (S_{\text{\tiny YM}}+S_{\text{\tiny{ghost}}}) } {\delta A_\mu}\right \rangle\,. 
\end{align}
While \eq{eq:nonsym} is trivial in Abelian theories (linear in the
expectation value of the field
$\sim \partial_\rho^2 F_{\mu\nu}(\langle A\rangle)\,$), it provides us
with a non-trivial loop relation in non-Abelian theories. Importantly,
in the tWTI there is no counter-part for the vanishing of the
Yang-Mills part in the longitudinal WTI in \eq{eq:nonsym} related to
gauge symmetry,
\begin{align}\label{eq:STIsym}
D_\mu \0{\delta S_{\text{\tiny YM}}}{\delta A_\mu} \equiv 0\,.
\end{align}  
This non-trivial relation is missing in the tWTI and signals that it is
not a gauge symmetry relation. 

Typically the tWTI is used within an Abelian approximation. Then
\eq{eq:nonsym} is trivial due to its independence of the quark fields
and it resembles the standard longitudinal WTIs or STIs in the same
approximation.

As has become clear from the discussion of the STIs,
the theory is fully non-perturbative  below
$\Lambda_{\text{\tiny{STI}}}$ and the Abelian approximation 
cannot hold any more. Strictly speaking one should even
loose confinement within the Abelian approximation. 
Hence, one should use the tWTI beyond the Abelian approximation 
in this regime. If evaluated for the quark-gluon vertex, the
ghost-contribution resembles the quark-ghost vertex in
the longitudinal STI, see \App{app:quarkghostscatteringkernel}. 

In conclusion, the Yang-Mills part in \eq{eq:nonsym} provides us with
an additional estimate for the STI scale as the scale where the
non-trivial Yang-Mills part neglected in Abelian approximations grows large. In this regime the
tWTI in the non-Abelian approximation provides us with an additional
functional relation for the quark-gluon vertex. Enforcing the tWTI
leads then to consistency between a solution of the quark-gluon vertex
DSE as derived from the functional quark and gluon DSE.\\

\section{Regulators}
\label{app:regulator}

For the regulators in \eq{eq:dSk}, we choose
\begin{align}
	& R_{k,\mu\nu}^{ab} &&= 
	\tilde{Z}_{A}(k)\, r(p^2/k^2)\, p^2\, \delta^{ab}\, 
\Pi^{\trans}_{\mu\nu}(p) \,,\\\nonumber
	& R^{ab}_k &&= Z_{c}(k)\, r(p^2/k^2)\, p^2 \, \delta^{ab} 
\,,\\\nonumber
	& R_k^q &&= Z_{q}(0)\, r(p^2/k^2) \, \slashed{p} \,,\\\nonumber
	& R_k^\phi &&= Z_{\phi}(0)\, r(p^2/k^2)\, p^2 \,,
	\label{eq:regulator}
\end{align}
where $r(p^2/k^2)$ denotes the dimensionless shape function 
and gluon regulator dressing $\tilde{Z}_{A}(k)$ is chosen as in 
\cite{Cyrol:2016tym}.
We employ the exponential regulator shape function, 
\begin{align}
	r(x)=\frac{x^{m-1}\,e^{-x^m}}{1-e^{-x^m}} \,.
\end{align}
In the pure glue system we have checked that the results obtained
with the exponential regulator agree, within our numerical precision, 
with the results obtained in \cite{Cyrol:2016tym} with the flat shape 
function.

\bibliographystyle{apsrev4-1}

\end{document}